\documentclass[showpacs,preprintnumbers,amsmath,amssymb]{revtex4}
\usepackage{graphicx}
\usepackage{dcolumn}
\usepackage{bm}
\usepackage{epsf}
\begin{document}
\newcommand{\beq}{\begin{equation}}
\newcommand{\eeq}{\end{equation}}
\newcommand{\half}{\frac{1}{2}}
\renewcommand{\d}{\partial}
\newcommand{\eq}[1]{eq.(\ref{#1})}
\newcommand{\la}[1]{\label{#1}}
\newcommand{\ur}[1]{(\ref{#1})}
\newcommand{\txt}[1]{{\mbox{\small #1}}}
\newcommand{\al}{\alpha}
\newcommand{\ah}{{\dot{\alpha}}/\alpha}
\newcommand{\ad}{\dot{\overline{\delta\alpha}}/\alpha}
\newcommand{\at}{\overline{\delta\alpha}/\alpha}
\newcommand{\de}{\delta}
\newcommand{\De}{\Delta}
\newcommand{\ga}{\gamma}
\newcommand{\Ga}{\Gamma}
\newcommand{\pa}{\partial}
\newcommand{\be}{\beta}
\newcommand{\si}{\simeq}
\newcommand{\sig}{\sigma}
\newcommand{\om}{\omega}
\newcommand{\ov}{\overline}
\newcommand{\Sm}{\mbox{}^{149}_{\:\:62}\text{Sm}}
\newcommand{\sr}{\hat{\sigma}_{r,Sm}(T_{C})}
\newcommand{\sre}{\hat{\sigma}_{r,Sm}(T_C,\Delta E_r)}
\newcommand{\U}{\mbox{}^{235}_{\:\:92}\text{U}}
\newcommand{\Pu}{\mbox{}^{239}_{\:\:94}{\rm Pu}}
\newcommand{\planck}{\text{\raisebox{.75\height}{--}\!\!\!{\it h}}}
\newcommand{\RZ}{{\em RZ2\ }}
\newcommand{\E}{\varepsilon}
\newcommand{\M}{{\cal M}}
\newcommand{\LT}{{\em LT\ }}
\title{Natural Nuclear Reactor Oklo and Variation of
Fundamental Constants: \\
Computation of Neutronics of Fresh Core
}
\author{Yu.V.Petrov}
\email{yupetrov@thd.pnpi.spb.ru}
\author{A.I. Nazarov}
\author{M.S. Onegin}
\author{V.Yu. Petrov}
\author{E.G. Sakhnovsky}
\affiliation{St. Petersburg Nuclear Physics Institute,
Gatchina, 188 300, St. Petersburg, Russia}
\date{June 8, 2005}
\begin{abstract}

Using  modern methods of reactor physics
we have performed full-scale calculations of
the natural reactor Oklo. For reliability we have used
recent version of two Monte Carlo codes:
Russian code MCU REA and world wide
known code MCNP (USA). Both codes produce similar results.
We have constructed a computer model of the reactor Oklo
zone \RZ
which takes into account all details
of design and composition. The calculations were performed for
three fresh cores with different uranium contents.
Multiplication factors, reactivities and neutron fluxes
were calculated.  We have estimated also the
{\em temperature and void effects} for the fresh core.
As would be expected, we have found for the fresh core a
significant difference between reactor and Maxwell spectra,
which was used before for averaging cross sections in the Oklo reactor.
The averaged cross section of $\Sm$ and its dependence
on the shift of resonance position (due to variation of
fundamental constants) are significantly different from previous
results.

Contrary to results of some previous papers we find no evidence
for the change of the fine structure constant in the past and
obtain new, most accurate limits on its variation with time:\\
$-4\cdot 10^{-17}$year
$\mbox{}^{-1}\leq \dot{\alpha}/\alpha \leq 3\cdot 10^{-17}$
year$\mbox{}^{-1}$. A further improvement in the accuracy of the limits
can be achieved by taking account of the core burnup. These calculations are
in progress.
\end{abstract}



\maketitle

\section{Introduction}

The discovery of the natural nuclear reactor in Gabon (West Africa) was
possibly one of the most momentous events in reactor physics since in
1942 Enrico Fermi with his team achieved an artificial self-sustained
fission chain reaction.
Soon after the discovery of the ancient natural reactor Oklo in Gabon
(West Africa) \cite{1,2,3}, one the authors of the present paper
(Yu.P.) and his postgraduate student A.I. Shlyakhter realized that
the "Oklo phenomenon" could be used to find the most precise limits on
possible changes of fundamental constants. At that time they considered
probabilistic predictions of unknown absorption cross sections based on
static nuclear properties. Near the neutron binding energy ($B_{n}=
6-8$ MeV) the resonances form a fence with a mean separation of tens
of electron volts. The magnitude of the cross section depends on the
proximity of the energy $E_{th}$ of the thermal neutron to the
nearest resonance.  If the energy $E_{th}=25$ meV falls directly on
a resonance, then the cross section increases to as much
$10^5-10^6$ b \cite{4a,4b,5}.  If the cross sections have changed with
time, then the entire fence of resonances as a whole has shifted by a
small amount $\Delta E_r$. This shift can be established most
accurately from the change of the cross section of strong absorbers
(for instance of $\Sm$).  The estimate of the shift $\Delta E_r$ on the
basis of experimental data for the Oklo reactor ($\Delta E_r \leq 3\cdot
10^{-17}$ eV/year) \cite{6a,6b,6c,7} allowed us to get the most
accurate estimate of a possible limit of the rate of change of
fundamental constants. This estimate has remained the  most accurate
one for 20 years. In 1996 a paper by Damour and Dyson was published in
which the authors checked and confirmed the results of Ref.\cite{6a,
6b, 6c}.  Damour and Dyson were the first to calculate the dependence
of the capture cross section on the temperature $T_C$ of the core:
$\sr$ \cite{8a,8b}.  In 2000 Fujii et al.  published a paper in which
the authors significantly reduced the experimental error of the cross
section $\sr$ \cite{9}.  In both papers the authors averaged the
samarium cross section with a Maxwell velocity spectrum over a wide
interval of the core temperature $T_C$.

After the publication of
Refs. \cite{8a, 8b}, one of the authors of the present paper (Yu. P.)
realized that the limit on the change of the cross section can be
significantly improved at least in two directions:
\begin{enumerate}
\item
Instead of a Maxwell distribution, the samarium cross section
should be averaged with the spectrum of the Oklo reactor which contains
the tail of Fermi spectrum of slowing down epithermal neutrons.
\item
The range of admissible core
temperatures $T_C$ can be significantly reduced, assuming that $T_C$ is
the equilibrium temperature at which the effective multiplication
factor $K_{eff}$ of the reactor is equal to one.
On account of the negative {\em power} coefficient
({\em void + temperature}) such a state of the
reactor will be maintained for a long time until the burn-up results in
a reduction of the reactivity excess and hence of $T_C$.
Since $\Sm$ burns
up about 100 times faster than $\U$, the core will contain
only that amount of samarium that was generated immediately before the
reactor shut down.  Therefore one needs to know the reactivity excess
and $T_C$ at the end of the cycle.
\end{enumerate}

To solve this problem one must use
modern neutron-physical and thermo-hydrodynamical methods of reactor
calculations. We
have built up a complete computer model of the Oklo reactor core \RZ
and established its material composition. We have chosen three
variants of its initial composition in order to estimate
its effect on the spread of results. To increase the
reliability of the results we have used modern versions of
two Monte Carlo codes. One of them, which has been developed at
the Kurchatov Institute, is the licensed Russian code MCU-REA
with the library DLC/MCUDAT-2.2 of nuclear data \cite{10};
the other one is the well known international code MCNP4C
with library ENDF/B-VI \cite{11}. Both codes give similar results.
We have calculated the multiplication
factors, reactivity and neutron flux for the fresh cores, and
the {\em void} and {\em  temperature effects} \cite{12}. As expected,
the reactor spectrum differs strongly from a Maxwell distribution (see
below). The cross section $\sre$, averaged with this
distribution, is significantly different from the cross section
averaged with a Maxwell distribution \cite{13}. We use our result for
the averaged cross section to estimate the position of resonances at
the time of Oklo reactor activity. This allows us to obtain the most
accurate limits on the change of the fine structure constant in the
past.

The paper is organized as follows. In section I we describe briefly
the history of the discovery of the natural Oklo reactor and itemize the
main parameters of its cores. We consider mainly the core \RZ. We
describe in detail the neutronics of this core calculated
by modern Monte Carlo codes. However simple semianalytical
considerations are also useful to clarify the picture. We consider
the {\em power effect} which is a sum of the {\em temperature} and
{\em void effects}. At the end of the section we discuss
the computational difficulties in the
calculations of the unusually large core \RZ
 and demonstrate that
Monte Carlo methods are, in general, inadequate for the calculations
of core burn-up.

The main result of Section I is the neutron spectrum in the fresh core.
In Section II we apply this spectrum to obtain the averaged cross
section of $\Sm$ in the past. We begin this Section with an
explanation of the way of obtaining precise limits on the
variation of fundamental constants using the available
Oklo reactor data. We describe different approaches to the problem and
relate the variation of the constants to the change in the averaged
cross sections for thermal neutrons. Using our value for the cross
section of $\Sm$ we obtain limits on the variation of the
fine structure constant which is the best available at the moment.
At the end of this Section we compare our result with the results
obtained in other papers and discuss possible reasons of differences.

\section{Neutronics of the fresh core}
\subsection{History of the discovery and parameters of the Oklo
reactor}
\subsubsection{History of the discovery of the natural reactor}

The first physicist to say in May or June of
1941 that a nuclear chain reaction could have been more easily realized
a billion years ago was Yakov Borisovich Zeldovich \cite{14}. At that
time he was considering the possibility of getting a fission chain
reaction in a homogeneous mixture of natural uranium with ordinary
water. His calculations (with Yu.B. Khariton) showed that this could be
achieved with an approximately two-fold enrichment of natural uranium
\cite{15, 16}.  A billion years ago the relative concentration of the
light uranium isotope was significantly higher, and a chain reaction was
possible in a mixture of natural uranium and water. ``Yakov Borisovich
said nothing about the possibility of a natural reactor, but his
thoughts directly lead us to the natural reactor discovered in Gabon in
1972'' reminisced I.I. Gurevich \cite{14}. Later, in 1957, G. Whetherill
and M. Inghram arrived at the same conclusion \cite{17a,17b}. Going
from the present concentration of uranium in pitchblende, they
concluded that about two billion years ago, when the proportion of $\U$
exceeded 3\%, conditions could be close to critical. Three years later,
P.  Kuroda \cite{18a,18b} showed that, if in the distant past there was
water present in such deposits, then the neutron multiplication factor
($K_\infty$) for an infinite medium could exceed unity and a
spontaneous chain reaction could arise.  But before 1972 no trace of a
natural reactor has been found.  On the 7th of June 1972, during a
routine mass-spectroscopic analysis in the French Pierrelatte factory that
produced enriched fuel, H.  Bouzigues \cite{1, 3} noticed that the
initial uranium hexafluoride contains $\zeta_5=0.717\%$ of $\U$ atoms
instead of the $0.720$\%, which is the usual concentration in
terrestrial rock, meteorites and lunar samples. The French Atomic
Energy Authority (CEA) began an investigation into this anomaly. The
phenomenon was named "Oklo phenomenon". The results of this research
were published in the proceedings of two IAEA symposia \cite{2,19}.
The simplest hypothesis of a contamination of the uranium by depleted
tails of the separation process was checked and shown to be wrong. Over
a large number of steps of the production process, the anomaly was
traced to the Munana factory near Franceville (Gabon) where the ore was
enriched. The ore with a mean uranium concentration of ($0.4-0.5$)\%
got delivered there from the Oklo deposit. The isotope analysis of the
uranium-rich samples showed a significant depletion of the $\U$ isotope
and also a departure from the natural distribution of those rare earth
isotopes, which are known as fission products \cite{1,3,20a,20b}. This
served as a proof of the existence in the distant past of a spontaneous
chain reaction. It had taken less than three months to produce this
proof.  A retrospective analysis of documents and samples of the Munane
enrichment factory showed that in 1970-72 ore was delivered for
processing that contained at times up to $20$\% of uranium depleted to
$0.64$\% of isotope $\U$ \cite{21}. Considering that the ore was mixed
during mining, the uranium concentration could be even higher in some
samples, and the depletion even stronger.  Altogether more than 700
tons of depleted uranium has been mined that had taken part in the
chain reaction. The deficit of $\U$ (that had not been noticed at
first) was about 200 kg. By agreement with the Government of Gabon, the
uranium ore production company of Franceville (COMUF) agreed to halt
mining in the region of the natural reactor. A Franco-Gabon group
headed by R.  Naudet began a systematic study of the Oklo phenomenon.
Numerous samples, obtained by boring, were sent for analysis to various
laboratories around the world. They allowed a reconstruction of the
functioning of the reactor in the Precambrian epoch.

\subsubsection{Geological history of the Oklo deposit}

As was shown by the $U/Pb$ analysis, the Oklo deposit with a
uranium concentration of about $0.5$\% in the sediment layer was
formed about $2\cdot 10^9$ years ago  \cite{22,23,24}. During this
epoch an important biological process was taking place: the transition
from prokaryotes, i.e. cells without nucleus, to more complex
unicellular forms containing a nucleus - eucaryotes. The eucaryotes
began to absorb carbon oxide and hence saturate the atmosphere with
oxygen. Under the influence of oxygen, the uranium oxides began
transforming into forms containing more oxygen, which are soluble in
water. Rains have washed them into an ancient river, forming in its
mouth a sandstone sediment, rich in uranium, of 4 to 10 meters
thickness and a width of 600 to 900 meters \cite{24}. The heavier
uranium particles settled more quickly to the ground of the nearly
stagnant water of the river delta. As a result the sandstone layer got
enriched with uranium up to 0.5\% (as in an enrichment factory). After
its formation, the uranium-rich layer, that was resting on a basalt
bed, was covered by sediments and sank to a depth of 4 kilometers. The
pressure on this layer was 100 MPa\cite{25}.  Under this pressure the
layer got fractured and ground water entered the clefts. Under the
action of the filtered water that was subjected to a high pressure, and
as a result of not completely understood processes, lenses formed with
a very high uranium concentration (up to $20-60$\% in the ore) with a
width of 10 to 20 meters and of the order of 1 meter thickness
\cite{26}.  The chain reaction took place in these lenses.  After the
end of the chain reaction the deposit was raised to the surface by
complicated tectonic processes and became accessible for mining. Within
tens of meters six centers of reactions were found immediately, and
altogether the remains of 17 cores were found \cite{27}.

The age $T_0$ of the reactor was
determined from the total number of $\U$ nuclei burnt up in the
past, $N_{5b}(d)$ , and the number of nuclei existing today,
$N_5(T_0)$(here $N_5$ is the density of $\U$ and $d$ is the duration of
the chain reaction).  For such a way of determining $T_0$ it is
necessary to know the number of $\Pu$ nuclei formed as a result of
neutron capture by $\mbox{}^{238}_{\:\:92}U$ and decayed to $\U$, and
the fluence $\Psi=\Phi d$ ($\Phi$ being the neutron flux).
Another independent method consists of the
determination of
the amount of lead formed as a result of the decay of $\U$, assuming
that it did not occur in such a quantity in the initial deposit
\cite{22}.  Both methods yield $T_0=1.81(5)\cdot 10^9$ years
\cite{7, 28}.
Below we assume in our calculations the value of $T_0=1.8\cdot
10^9$ years.

The duration of the work of the reactor can be established
from the amount of $\Pu$ formed. One can separate the decayed
$\Pu$ from the decayed $\U$ using the different relative yields of
$\text{Nd}$ isotopes: \\ $\delta^9_{Nd}$
$=\mbox{}^{150}\text{Nd}/(\mbox{}^{143}\text{Nd}+
\mbox{}^{144}\text{Nd})=0.1175$ for $\Pu$ and $\delta^9_{Nd}=0.0566$ for
$\U$ \cite{29}. However this comparison is
masked by the fission of $\mbox{}^{238}_{\:\:92}\text{U}$ by fast neutrons:
$\delta^8_{Nd}=0.1336$.  Taking account of this contribution one
arrives at an estimate of $d\sim 0.6$ million years \cite{30}. This was
the value we adopted in our calculations.

The total energy yield of the reactor
has been estimated to be $1.5\cdot 10^4$ MWa \cite{31}. Such a fission
energy is obtained by two blocks of the Leningrad atomic power station
with a hundred percent load in 2.3 years. Assuming a mean duration of $d
= 6\cdot 10^5$ a for the work of the reactor one gets a mean power
output of only $P_P =25$ kW.

\subsection{Composition and size of the Oklo \RZ reactor}

The cores of the Oklo
reactor have been numbered. The most complete data are available for
core \RZ.  This core of the Oklo reactor is of the shape of an
irregular rectangular plate that lies on a basalt bed at an angle of
45$^o$. The thickness of the plate is $H=1$ m, its width is $b=11-12$
m, and its length is $l=19-20$ m (see Fig.8a in Refs.\cite{31} and
\cite{22}).  Thus the volume of the \RZ core is about 240 m$^3$.
Since in the case of large longitudinal and transverse sizes the shape
of the reactor is not essential, we have assumed as a reactor model a
flat cylinder of height $H=1$ m and radius $R$ which is determined by
the core burn-up. The energy yield is $P_Pd=1.5\cdot10^4$ MWa $\si
5.48\cdot10^6$ MWd. At a consumption of $\U$ of $g=1.3$ g/MWd\cite{32},
the total amount of burnt up fissile matter is
\beq
\De M_b=gP_Pd=7.12 \, {\rm tons}.
\eeq
Taking into
account that half of the burnt up $\U$ isotope is replenished from the
decay of the produced $\Pu$, we find the original mass of the
burnt up $\U$:
\beq
\De M_5=4.75 \, {\rm tons}.
\eeq
In case of a uniform burn-up, the average
density of the burnt up $\U$ is
\beq
\De\ga_5(d)=\frac{\De M_5}{\pi R^2H}=\frac{1.51}{R^2} \, {\rm g/cm}^3,
\eeq
where $R$ is given in meters.  The relative average initial
burn-up is
\beq \ov{y}_5(d)=\frac{\De\ga_5(d)}{\ga_{5,i}(0)}=
\frac{1.51 \,{\rm g/cm^3}}{\ga_{5,i}(0)R^2_i} \quad {\rm and} \quad
R_i =\Biggl[\frac{1.51 \, {\rm
g/cm^3}}{\ga_{5,i}(0)\ov{y}_5(d)}\Biggr]^{1/2}.
\eeq
Processing the
data of Table 2 from Ref.\cite{31} gives a value of $\ov{y}_5(T_0)\si
50\%$ for the present-day average over the core. In the past it was
1.355 times smaller (see below) on account of the
higher concentration $\ga_{5,i}(0)$ of uranium, i.e.
$\ov{y}_5(0)=36.9\%$.  Thus the radius is given by the following
formula:
\beq
R_i=\Biggl[\frac{4.09 \,{\rm
g/cm^3}}{\ga_{5, i}(0)}\Biggr]^{1/2} \,{\rm m}
\la{eq4a}
\eeq

The approximate composition of the rock in core \RZ is shown in
Table 1 of Ref.\cite{7}.  On the basis of these data one can calculate
the elemental composition of the ore by weight (see the penultimate
column of Table\ref{tab1}). For comparison we show in the last column
the composition by weight from the book of Yu. A.  Shukolyukov (Table
2.1 of Ref.  \cite{33}), which was based on early data of R. Naudet
\cite{34}.  These values coincide within 10$\%$. Since all those
elements are relatively weakly absorbing, such differences
practically do not play any role.

\begin{table}[ht]
\caption{\la{tab1}
Present-day composition of the empty rock \cite{7}}
\begin{tabular}{|c|c|c|c|c|c|c|}
\cline{3-7}
\multicolumn{2}{c|}{} &{\em 1} &{\em 2} &{\em 3} &{\em 4} &{\em 5} \\
\hline
&  Chemical   &\%  by & Elemental &Atomic  & \% by & \% by \\
&  composition&weight  &composition&weight, $A_i$ & weight &
weight \cite{33} \\
\hline
1& SiO$_2$ & 43.00& O & 15.999& 44.04 & 44 \\
\hline
2& Al$_2$O$_3$ &25.73 & Si & 28.086 & 20.10 & 20\\
\hline
3& FeO & 14.53 & Al & 26.982& 13.62& 16 \\
\hline
4&  Fe$_2$O$_3$& 4.47& Fe & 55.847& 14.42 & 11\\
\hline
5 & MgO & 10.43 & Mg  & 24.305 & 6.30 & 4\\
\hline
6 & K$_2$O & 1.84 & K & 39.098 & 1.53 & 2\\
\hline
7& Sum & 100 & &&     100& 97 \\
\hline
\end{tabular}
\end{table}

\begin{table}[hb]
\caption{\la{tab2}
Present-day and initial composition of the ore in core \RZ \cite{35a}}
\begin{tabular}{|c|l|c|c|c|c|} \cline{3-6}
\multicolumn{2}{c|}{} & $i$ & {\em 1} &{\em 2} &{\em 3}  \\
\hline
1& Present-day fraction of U in the & $Y_{U,i}(T_0)$ & 35& 45 &55 \\
 & dry ore, \%    &&&& \\
\hline
2& Present-day density of the dry & $\ga_i(T_0)$ & 2.84& 3.29 & 3.82 \\
 & ore, g/cm$^3$ (Fig.\ref{fig1}) &&&& \\
\hline
3& Present-day density of U in the& $\ga_{U,i}(T_0)$& 0.994& 1.481&
2.101  \\
 & dry ore, g/cm$^3$ &&&& \\
\hline
4& Present-day density of UO$_2$ in& $\ga_{UO_2,i}(T_0)$& 1.128& 1.680&
2.38 \\
 & the dry ore, g/cm$^3$ &&&& \\
\hline
5& Present-day density of the dry& $\ga_{0,i}(T_0)$& 1.71& 1.61& 1.44 \\
 & rock with Pb, g/cm$^3$ &&&& \\
\hline
6& Density of H$_2$O at T=300K& $\ga_{H_2O}$& 0.355& 0.355& 0.355\\
 & (with water of crystallization)&& 0.455& 0.455&  0.455 \\
\hline
7& Initial density of $^{238}_{92}$U, g/cm$^3$& $\ga_{8,i}(0)$&  1.305&
  1.952&   2.758 \\
\hline
8&Initial density of $^{235}_{92}$U, g/cm$^3$& $\ga_{5,i}(0)$&
   4.216$\cdot$10$^{-2}$& 6.28$\cdot$10$^{-2}$& 8.91$\cdot$10$^{-2}$ \\
\hline
9& Initial density of U, g/cm$^3$ &$\ga_{U,i}(0)$&1.347&2.015&2.847 \\
\hline
10&Initial density of UO$_2$, g/cm$^3$& $\ga_{UO_2,i}(0)$& 1.528&
  2.286&   3.230 \\
\hline
11& Initial density of ore, g/cm$^3$& $\ga_i(0)$& 3.51& 4.09& 4.78 \\
\hline
12& Initial density of ore without& $\ga_{0,i}(0)$& 1.98& 1.80& 1.55 \\
  & UO$_2$, g/cm$^3$ &&&& \\
\hline
13& Initial fraction of U in the& $Y_{U,i}(0)$& 38.4& 49.3&
59.6 \\
  & dry ore, \%  &&&& \\
\hline
\end{tabular}
\end{table}

It is  much more important to know the amount of uranium and of water
at the beginning of the work of the reactor. The connection between the
uranium content in the core (in $\%$) and the density of the dehydrated
core has been measured experimentally in Ref.\cite{35}
(Fig.\ref{fig1}).  The content of uranium in the core varies greatly
between different samples. To determine the influence of the uranium
content on the reactor parameters we have chosen three initial values
for the density of uranium in the dehydrated ore: $Y_{U,i}(T_0)=$35, 45
and 55\%, taken to be constant over the reactor. The value of the ore
density $\gamma_i(T_0)$ that corresponds to $Y_{U,i}(T_0)$ is shown
in the second row of Table \ref{tab2} (see Fig.\ref{fig1}).  In the
fifth row of Table \ref{tab2} we show the density of the empty rock.
The density of water in the reactor is $0.3-0.5$ g/cm$^3$ \cite{27}.
This water consists of bound (crystalline) and unbound water which
evaporates after 100$^o$C.  In our reactor model we assumed a total
density of water of $\ga_{H_2O}=0.355$ g/cm$^3$, of which 0.155
g/cm$^3$ was taken for the density of unbound water.  Assuming a
porosity of about 6\% one can take for the density of dry ore with
water the same value as for the dehydrate ore.

\begin{figure}[h]
\centerline{
\epsfbox{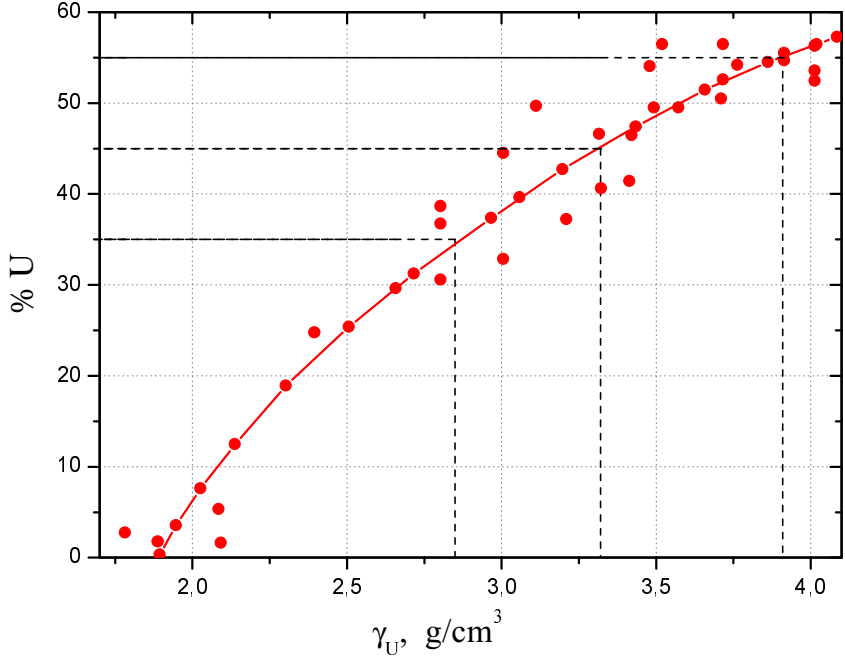}}
{\caption{\la{fig1}
Present-day abundance of  uranium (in volume percent of dry ore)
in dependence on the dry ore density $\ga_{\rm U}$ in the Oklo reactor
\cite{35}. Three initial variants of the $Y_{U, i}(T_0)=35\%,45\%,55\%$U
in dry ore are shown.}} \end{figure}

The density of $\mbox{}^{238}_{\:\:92}$U of the fresh core \RZ at the
epoch of the formation of the reactor was
\beq
\ga_8(0)=\ga_8(T_0)(1-\zeta_5){\rm exp}(+T_0/\tau_8),
\la{eq5}
\eeq
 where the lifetime of
$\mbox{}^{238}_{\:\:92}$U is $\tau_8=6.45\cdot10^9$ year. The value of
$\ga_U^8(0)$ increases on account of the decay of uranium into lead.
The density of $\U$ ($\ga_{U,5}(0)$, g/cm$^3$) in the fresh core
($\tau_5=1.015\cdot10^9$ year) is
\beq
\ga_{U,5}(0)=\ga_{U,5}(T_0)\zeta_5{\rm exp}(+T_0/\tau_5).
\la{eq6}
\eeq
The values of $\ga_{5,i}(0)$ and $\ga_{8,i}(0)$ are shown in
Table \ref{tab2}.  Also in the table are the calculated values of the
densities of uranium $\ga_{U,i}(0)$ and of the empty rock (without Pb),
and the new fraction of uranium in the dry ore at the beginning of the
cycle of the Oklo reactor. From \eq{eq5} and \eq{eq6} we get for the
ratio $\ga_{U,i}(0)/\ga_{U,i}(T_0)$ \beq
\ga_{U,i}(0)/\ga_{U,i}(T_0)=(1-\zeta_5){\rm exp}(T_0/\tau_8)+
\zeta_5{\rm exp}(T_0/\tau_5)=1.355,
\la{eq7}
\eeq
independent of $\ga_{U,i}(T_0)$. The initial concentration
$N_{k,i}(0)$ of nuclei, which is needed for the calculations, was
calculated from the formula
\beq
N_{k,i}(0)=\ga_{k,i}(0)N_A/A_k,
\eeq
where $N_A=6.022\cdot10^{23}$ mol$^{-1}$  is the Avogadro number and
$A_k$ is the atomic weight.  For $\mbox{}^{238}_{\:\:92}$U and $\U$ we
have used the data of Table \ref{tab2}; for the atoms of the rock we
used the percentages by weight from Table \ref{tab1}. The oxygen
content of water was added to the oxygen of the core. The composition
of the fresh core \RZ that we used in calculations with different
initial content of uranium is shown in Table \ref{tab3}. Although the
accuracy of the densities of some elements in this table is only a few
percent, the values of $N_{k,i}(0)$ are given with four decimal places
for reproducibility of results.

\begin{table}[!]
\caption{\la{tab3}
Specific weight $\ga_{k,i}(0)$ and nuclear density $N_{k,i}(0)$
in the compositions of three variants of the reactor core \cite{35a}}
\begin{tabular}{|c|c|c|c|c|c|c|c|c|} \cline{3-9}
\multicolumn{2}{c|}{}& $i$& \multicolumn{2}{|c|}{\it 1}&
\multicolumn{2}{|c|}{\it 2}& \multicolumn{2}{|c|}{\it 3} \\
\cline{3-9}
\multicolumn{2}{c|}{}& $Y_{U,i}(0)$, $\%$& \multicolumn{2}{|c|}{38.4}&
\multicolumn{2}{|c|}{49.4}& \multicolumn{2}{|c|}{59.6} \\
\hline
$k$& Elemental& Atomic weight,& $\ga_{k,1}(0)$& $N_{k,1}(0)$&
$\ga_{k,2}(0)$& $N_{k,2}(0)$& $\ga_{k,3}(0)$ & $N_{k,3}(0)$ \\
   & composition& $A_K$, g/mol& g/cm$^3$& $(b\cdot cm)^{-1}$&
  g/cm$^3$& $(b\cdot cm)^{-1}$& g/cm$^3$& $(b\cdot cm)^{-1}$ \\
\hline
1& $\U$& 235.04& 4.216$\cdot10^{-2}$& 1.0802$\cdot10^{-4}$&
  6.28$\cdot10^{-2}$& 1.6090$\cdot10^{-4}$& 8.91$\cdot10^{-2}$&
  2.2828$\cdot10^{-4}$ \\
\hline
2& $^{238}_{\:\:92}$U& 238.05& 1.305& 3.3013$\cdot10^{-3}$& 1.952&
   4.9380$\cdot10^{-3}$& 2.758& 6.9770$\cdot10^{-3}$ \\
\hline
3& $_{92}$U&       & 1.347& 3.4093$\cdot10^{-3}$& 2.015&
   5.0989$\cdot10^{-3}$& 2.847& 7.2052$\cdot10^{-3}$ \\
\hline
4& $_8^{16}$O& 15.999& 1.3683& 5.1510$\cdot10^{-2}$& 1.3790&
   5.1906$\cdot10^{-2}$& 1.3809& 5.1978$\cdot10^{-2}$ \\
\hline
5& $^1_1$H& 1.0079& 3.97$\cdot10^{-2}$& 2.373$\cdot10^{-2}$&
  3.97$\cdot10^{-2}$& 2.373$\cdot10^{-2}$& 3.97$\cdot10^{-2}$&
  2.373$\cdot10^{-2}$ \\
\hline
6& $_{14}$Si& 28.086& 0.398& 8.534$\cdot10^{-3}$& 0.362&
   7.762$\cdot10^{-3}$& 0.312& 6.690$\cdot10^{-3}$ \\
\hline
7& $_{13}^{27}$Al& 26.982& 0.270& 6.026$\cdot10^{-3}$& 0.245&
   5.468$\cdot10^{-3}$& 0.211& 4.709$\cdot10^{-3}$ \\
\hline
8& $_{12}$Mg& 24.305& 0.125& 3.097$\cdot10^{-3}$& 0.113&
   2.800$\cdot10^{-3}$& 0.097& 2.403$\cdot10^{-3}$ \\
\hline
9& $_{26}$Fe&  55.847& 0.286& 3.084$\cdot10^{-3}$& 0.260&
   2.804$\cdot10^{-3}$& 0.224& 2.415$\cdot10^{-3}$ \\
\hline
10& $_{19}$K&  39.098& 0.0302& 4.652$\cdot10^{-4}$& 0.0275&
   4.235$\cdot10^{-4}$& 0.0237& 3.653$\cdot10^{-4}$ \\
\hline
\multicolumn{2}{c|}{}& $\sum_{k=3}^{10}\ga_{k,i}$&
    3.509& & 4.081& & 4.780 & \\
\cline{3-9}
\end{tabular}
\end{table}

It follows from Table \ref{tab2} that the enrichment of isotope $\U$
($\zeta_5(0)=\ga_5(0)/\ga_U(0)$) was $\zeta_5(0)=3.1\%$ 1.8 billion
years ago.  Uranium of such enrichment is used in Russian VVER reactors
of atomic power stations.  Since the ratio of nuclei U/H is about equal
and the sizes of both reactors are comparable, one can immediately and
without any calculation say that a chain reaction was possible in Oklo
\cite{7}.

\subsection{Calculation of the fresh core}

\subsubsection{Semianalytical
calculation of core \RZ}

Consider first the
bare reactor without reflector. Since the reactor is large compared
with the neutron migration length $M$, one can apply the single-group
diffusion theory \cite{Wein,32}. For a stationary neutron flux
$\Phi(\vec r)$  the following equation holds:
 \beq
\Biggl[-\nabla^2+\frac{1}{M^2}\Biggr]
\Phi(\vec r)=\frac{K_\infty}{K_{\rm eff} M^2}\,\Phi(\vec r), \quad
\Phi\left(\pm\frac H2\right)=\Phi(R)=0\,.
\la{eq9}
\eeq
The solution $\Phi(\vec r)$ that satisfies
this equation with boundary conditions \eq{eq9} is
\beq
\Phi(\vec r)=\Phi_0{\rm cos} \biggl(\frac{\pi}{H} x\biggr) J_0
\biggl(\frac{2,405 r}{R}\biggr)\, ,
\eeq
where $J_0(B r)$  is the
zeroth Bessel function, and the effective multiplication factor is
\beq
K_{\rm eff}=\frac{K_\infty}{1+M^2B^2}\, ; \quad B^2=B^2_H+B^2_R\, ;
\quad B_H=\frac{\pi}{H}\, ; \quad B_R=\frac{2.405}{R}\, .
\la{eq11}
\eeq
In Table \ref{tab4} we show the two constants, $K_{\infty, i}$ and
$M^2_i$, calculated with codes MCNP4C and MCU-REA for three different
cores \cite{35a,35b}.  These constants are needed to calculate $K_{{\rm
eff}, i}$ by formula \ur{eq11}.  The values of $K_{\infty, i}$
calculated for one and the same composition differ by a few tenth of a
per cent; the values of $M^2_i=K_{\infty, i}\tau_i+L^2_i$ differ by a
few per cent.  In row 9 of Table \ref{tab4} we show the values of
$K_{{\rm eff}, i}^{(1)}$ calculated with the approximate formula
\ur{eq11}.  They are smaller than the direct calculations using Monte
Carlo code (row 1).  The difference in reactivity amounts to
$\De\rho_1=-(0.2-0.3)\%$.  The diffusion length in the fresh core is
$L=1.6-2.1$ cm, and the total migration length is $M=6-7$ cm.  These
lengths get less with increasing uranium concentration.

\begin{table}[!]
\caption{\la{tab4}
Two-group parameters of the fresh cores of a cylindrical bare
reactor with different content of uranium $i$. Thickness of the core
$H=1$m; average temperature in the core $T=300$K; density of water
$\ga_{H_2O}=0.355$ g/cm$^3$ \cite{35a}}

\begin{tabular}{|c|l|c|c|c|c|c|c|} \hline
\multicolumn{2}{|l|}{Variant of the core}&
\multicolumn{2}{|c|}{\it 1}& \multicolumn{2}{|c|}{\it 2}&
\multicolumn{2}{|c|}{\it 3} \\ \hline
\multicolumn{2}{|l|}{Relative initial density of U}& \multicolumn{2}{|c|}{}&
\multicolumn{2}{|c|}{}& \multicolumn{2}{|c|}{} \\
\multicolumn{2}{|l|}{in the ore $Y_{U,i}(0)$, $\%$}&
\multicolumn{2}{|c|}{38.4}& \multicolumn{2}{|c|}{49.42}&
\multicolumn{2}{|c|}{59.6} \\
\hline
\multicolumn{2}{|l|}{Radius of the active core, $R$, m }&
\multicolumn{2}{|c|}{ 9.9}& \multicolumn{2}{|c|}{ 8.1}&
\multicolumn{2}{|c|}{ 6.8} \\ \hline
\multicolumn{2}{|l|}{Computer code}& MCNP4C& MCU REA&  MCNP4C&
MCU REA& MCNP4C& MCU REA     \\ \hline
1& $K_{eff}$& 1.0965(1)& 1.0971(1)& 1.1238(2)& 1.1271(1)&
   1.1247(2)& 1.1306(1)  \\ \hline
2& $K_\infty$& 1.1501(2)& 1.1499(2)& 1.1750(2)& 1.1771(1)&
   1.1721(1)& 1.1769(1)  \\ \hline
3& Leakage of fast (${\cal L}_F$), & & & & & & \\
 & ($E>0.625$ eV)& 4.02$\cdot10^{-2}$&
   3.96$\cdot10^{-2}$& 3.84$\cdot10^{-2}$& 3.76$\cdot10^{-2}$&
   3.63(1)$\cdot10^{-2}$&    3.55$\cdot10^{-2}$ \\
\hline
4& Leakage of thermal (${\cal L}_{th}$), & & & & & &    \\
 & ($E<0.625$ eV)& 0.434$\cdot10^{-2}$&
   0.435$\cdot10^{-2}$& 0.317$\cdot10^{-2}$& 0.316$\cdot10^{-2}$&
   0.230(1)$\cdot10^{-2}$& 0.230$\cdot10^{-2}$   \\
\hline
5& Square of diffusion length   & & & & & &  \\
 & $L^2={\cal L}_{th}/[(1-{\cal L}_{th})B^2_{00}]$ cm$^2$&
4.4& 4.4& 3.2& 3.2& 2.3& 2.3 \\ \hline 6& Age
$\tau=-\log(1-{\cal L}_F)/B^2_{00}$, cm$^2$& 41.3& 40.7& 39.3& 38.5&
37.0& 36.1 \\ \hline 7& Total migration area       & & & & & & \\ &
$M^2=K_\infty\tau+L^2$, cm$^2$& 51.9& 50.2& 49.4& 48.5& 45.8& 44.8 \\
 \hline 8& $B^2_{00}=\pi^2/H^2+2.4048^2/R^2$,
cm$^{-1}$; & 0.99292$\cdot10^{-3}$& 0.99584$\cdot10^{-3}$&
   0.99954$\cdot10^{-3}$& && \\ &
 $\pi^2/H^2=0.98696\cdot10^{-3}$, cm$^{-2}$& & & & && \\ \hline
9& $K_{eff}^{(1)}$ by Eq.\ur{eq11}& 1.0937(2)& 1.0943(2)& 1.1199(2)&
   1.1228(1)& 1.1208(2)& 1.1265(1)   \\ \hline
10& Account of Eq.\ur{eq11} $\De\rho_{ap}-\rho_{eff}$, $\%$& $-$0.23(2)&
   $-$0.23(2)& $-$0.31(2)& $-$0.34(2)& $-$0.31(3)& $-$0.32(1) \\
\hline
\end{tabular}
\end{table}

The mean neutron flux, averaged over the reactor, is
\beq
\ov\Phi=\frac{1}{V}\int_V \Phi(\vec r)d\vec r= \Phi_0\frac{4}{\pi}
\frac{J_1(2.405)}{2.405}
\la{eq12}
\eeq
($J_1(2.405)=0.51905$).  From formula \ur{eq12} we get the following
formula of the volume nonuniformity coefficient $K_V$, independent of
$R$ and $H$:
\beq
K_V=\Phi_0\bigl/\ov\Phi=\frac{\pi}{2}\cdot\frac{2.405}{2J_1(2.405)}=
3.638\, .
\la{eq12a}
\eeq
This  formula is useful to check the accuracy of calculation of the
spatial distribution $\Phi(\vec r)$.  The absolute value of the mean
neutron flux for $P_P=2.5\cdot10^{-2}$ MW is equal to \beq
\ov\Phi=\varphi\frac{v_f}{E_f}P_P=1.88\cdot10^{15} \,\, {\rm
n/s}\cdot\varphi\, ,
\la{eq13}
\eeq
where $\varphi$ is the
neutron flux per cm$^2$ and one fast fission neutron which is calculated
with the Monte Carlo code; $\nu_f/E_f=7.5\cdot10^{16}$ n/MW$\cdot$s is
the number of fast neutrons per second and a power of 1 MW ($E_f$
is the fission energy; $\nu_f$ is the number of fast neutrons per
fission).  For thermal neutrons formula \ur{eq13} holds with
$\varphi_{\rm th}$.  The mean thermal neutron flux with energies
$E_n<0.625$ eV is very small in the case of $Y_{U,2}(0)=49.4\%$ it is
$\ov\Phi_{\rm th}=0.63\cdot10^8$ n/cm$^2\cdot$s.  The thermal flux in
the center of the core is $\Phi_{\rm th}^{\rm max}(0)=2.00\cdot10^8$
n/cm$^2$s.  The total mean flux, integrated over all energies, is equal
to $\ov\Phi=3.9\cdot10^8$ n/cm$^2$s.  These results were found using
code MCU-REA.  The results of calculations using other Monte Carlo
codes are similar (see Table \ref{tab5}).  The low neutron flux
determines the specifics of the function of the reactor.

\begin{table}[ht]
\caption{\la{tab5}
Calculations of the total flux $\ov{\Phi}_{tot}$  averaged over
the bare reactor,
and of the average thermal flux \\$\ov{\Phi}_{th}$ ($E_n<0.625$ eV) for
the fresh core using three different codes. $Y_{02}(0)=49.4\% U$ in the
dry ore; $\omega_{H_2O}^0=0.355$;\\ $P_C=0.1$ MPa; $T=300$K;
$P_P=25$~kW}

\begin{tabular}{|l|c|c|c|} \cline{2-4}
\multicolumn{1}{c|}{}&  MCU-REA & MCB & MCNP4C  \\
\hline
$K_{eff}$ &   1.1271(1)& 1.1237(1) &  1.1243(2) \\
\hline
$\ov{\Phi}_{tot}$, \, $\cdot10^8$ n/cm$^2$s& 3.9& 4.4 & 4.4 \\
\hline
$\ov{\Phi}_{th}$\, ($E_n<0.625$ eV), $\cdot10^8$ n/cm$^2$s&
      0.63&        0.64&       0.63 \\
\hline
\end{tabular}
\end{table}

As a
reflector one can assume the same core but without uranium. The
analytical calculations for the reactor with reflector are more
cumbersome.  Therefore we used numerical methods for these
calculations.  The results are shown in Table \ref{tab6}. Both Monte
Carlo programs give values of the reactivity reserve for variant {\em
1} of the core which coincide within the statistical accuracy. The
difference of reactivity is 0.26\% for core \RZ of the bare reactor and
0.46\% for core {\em 3}. For the reactor with reflector the difference
is smaller:  0.20\% and 0.39\%, respectively.  Since the migration
length is small, a reflector of thickness $\De=0.5$ m  is practically
infinite:  the results for a reflector of thickness $\De=0.5$ m
coincide with those for $\De=1$ m.  Compared with the bare reactor, the
reflector makes a contribution of $\delta\rho(\De)=0.8\div0.9\%$.
This contribution drops with increasing uranium content in the core.
The cold reactor with a fresh core is strongly overcritical, since
{\em temperature} and {\em void} effects have not yet been taken into
account, also the initial strong absorbers which afterwards burn up
rapidly. In Table \ref{tab7} we show the neutron capture in the
infinite fresh core per fast fission neutron.  Capture by $\U$ amounts
to 55.7\% and by $\mbox{}^{238}_{\:\:92}$U to 33.8\%.  These are
followed by hydrogen (3.9\%), iron (3.8\%), silicon (0.8\%) etc.  Code
MCNP4C gives similar values.

\begin{table}[h]
\caption{\la{tab6}Calculation of $K_{eff}$ and $\rho_i$ for
cylindrical core of thickness $H_0=1$m and radius $R$ for three
different initial contents of uranium in the dry ore. Density of
water $\ga_{H_2O}=0.355$g/cm$^3$; $T=300$K \cite{35a}}

\begin{tabular}{|c|l|c|c|c|c|c|c|c|c|c|} \hline
\multicolumn{2}{|c|}{Variant of core}& \multicolumn{3}{|c|}{\it 1}&
\multicolumn{3}{|c|}{\it 2}&
\multicolumn{3}{|c|}{\it 3} \\
\hline
 & Relative initial density of U in & \multicolumn{3}{|c|}{}&
\multicolumn{3}{|c|}{}& \multicolumn{3}{|c|}{} \\
& core $Y_{U,i}(0)$, \%& \multicolumn{3}{|c|}{38.4}&
\multicolumn{3}{|c|}{ 49.4}& \multicolumn{3}{|c|}{ 59.6} \\
\cline{2-11}
 & Specific weight of the dry ore&   \multicolumn{3}{|c|}{}&
\multicolumn{3}{|c|}{}& \multicolumn{3}{|c|}{} \\
1& $\gamma_i(0)$, g/cm$^3$& \multicolumn{3}{|c|}{3.51}&
\multicolumn{3}{|c|}{ 4.08}& \multicolumn{3}{|c|}{ 4.78} \\
\cline{2-11}
 & Radius of active core $R$, m& \multicolumn{3}{|c|}{9.85}&
\multicolumn{3}{|c|}{ 8.07}& \multicolumn{3}{|c|}{ 6.78} \\
\cline{2-11}
 & Thickness of reflector $\De$, m& 0& 0.5& 1& 0& 0.5& 1&
 0& 0.5& 1 \\ \hline
& MCNP4C:  & \multicolumn{3}{|c|}{}& \multicolumn{3}{|c|}{}&
\multicolumn{3}{|c|}{} \\
 & $K_{eff}$& 1.0965(1)& 1.1077(1)& 1.1077(1)& 1.1238(2)& 1.1357(2)&
 1.1352(1)& 1.1247(2)& 1.1353(2)& 1.1357(1)     \\
2& Reactivity $\rho_i$, $\%$& 8.80(1)& 9.72(1)& 9.72(1)& 11.02(2)&
   11.95(2)& 11.91(1)& 11.09(2)& 11.92(2)& 11.95(1)   \\
 & $\delta\rho_i(\De)=\rho_i(\De)-\rho_i(0)$& 0& 0.92(1)& 0.92(1)&  0&
   0.93(3)& 0.89(2)& 0& 0.83(3)& 0.86(1)  \\ \hline
 & MCU-REA:  & \multicolumn{3}{|c|}{}& \multicolumn{3}{|c|}{} &
\multicolumn{3}{|c|}{} \\
 & $K_{eff}$& 1.0971(1)& 1.1080(1)& 1.1080(1)& 1.1271(1)& 1.1383(1)&
   1.1381(1)& 1.1306(1)& 1.1407(1)& 1.1405(1) \\
3& Reactivity $\rho_i$, $\%$& 8.85(1)&  9.75(1)& 9.75(1)& 11.28(1)&
   12.15(1)& 12.13(1)& 11.55(1)&  12.33(1)&  12.32(1) \\
 & $\delta\rho_i(\De)=\rho_i(\De)-\rho_i(0)$& 0& 0.90(1)& 0.90(1)& 0&
  0.87(1)& 0.85(1)& 0&  0.78(1)& 0.77(1) \\
\hline
\end{tabular}
\end{table}

\begin{table}[h]
\caption{\la{tab7}
Absorption $<\Sigma_a\Phi V>_k$ and absorption with fission
$<\Sigma_f\Phi V>_k$ for infinite medium, normalized to one capture, for
core variant $Y_{U,i}(0)=49.4\% U$; $T=300$K; $\%$}

\begin{tabular}{|c|l|c|c|c|c|} \cline{3-6}
\multicolumn{2}{c|}{} & \multicolumn{2}{|c|}{MCNP4C}&
  \multicolumn{2}{|c|}{MCU REA}  \\ \cline{3-6}
\multicolumn{2}{c|}{}& $<\Sigma_a\Phi V>$& $<\Sigma_f\Phi V>$&
$<\Sigma_a\Phi V>$& $<\Sigma_f\Phi V>$   \\ \hline
1& $^{235}$U& 55.6& 45.9& 55.7& 46.0 \\ \hline
2& $^{238}$U& 34.0&  2.1& 33.8&  2.1 \\ \hline
3& O        &  0.7&     &  0.7&       \\ \hline
4& H        &  4.0&     & 3.9&        \\ \hline
5& Si       &  0.8&     &  0.8&       \\ \hline
6& Al       &  0.7&     &  0.7&      \\
\hline
7& Mg       &  0.1&     &  0.1&      \\
\hline
8& Fe       &  3.7&     &  3.8&     \\
\hline
9& K        &  0.5&     &  0.5&     \\
\hline
10& $\sum^9_{K=1}(\Sigma_a\Phi V)_K$& 100.0& 48.0& 100.0& 48.1 \\
\hline
\end{tabular}
\end{table}

\subsubsection{Power effect}

Reactor Oklo is controlled by the core
temperature $T_C$ \cite{36}.  During heating the water was driven out
of the core until the multiplication factor was equal to one. At first
the large overcriticality was compensated by the {\em power effect},
which is the sum of the {\em temperature} and {\em void effects}. In
Table \ref{tab8} and Fig.\ref{fig2} we show the dependence of the water
density on the temperature for several pressures. At a pressure of 100
MPa in the Oklo reactor and $T_C=700$K, the density of water is 65\% of
its value for $T_C=300$K and normal pressure. In this case the
difference between crystalline and free water disappears apparently.
The {\em power effects} is shown in Fig.\ref{fig3} \cite{35a,35b}.
Near $T_C=700$K all the $K_{{\rm eff} i}$  become equal to one.
Therefore we can assume $T_C\si 700$K as the most likely temperature of
the fresh active core (we neglect the small difference between the
temperatures of fuel and water).  For the variant of the composition of
core \RZ, the {\em power effect} is $\De\rho_P=-11.6\%$. The {\em void
effect} at 700 K accounts for 73\% of this value, and the {\em
temperature effect} for 27\%.  These results were obtained with code
MCNP4C. Code MCU-REA gives similar values. In Table \ref{tab4} we show
the numerical values of $K_{\rm eff}(T_C)$, $K_\infty(T_C)$  and
$M^2(T_C)$, calculated with code MCU-REA for the bare reactor with core
composition $i=2$. With increased temperature $K_\infty(T_C)$ drops and
$M^2$ and the leakage increase (Fig.\ref{fig4}).

\begin{table}[h]
\caption{\la{tab8}
Temperature dependence of the water density at a pressure of
$P_C=100$ MPa in the core}

\begin{tabular}{|c|c|c|c|} \hline
$T(K)$& $\ga_{H_2O},$ g/cm$^3$ & $\om_1\ga_{H_2O},$ g/cm$^3$ &
$\om_2\ga_{H_2O},$ g/cm$^3$ $^{1)}$  \\ \hline
300&  1.037& 0.368&  0.472 \\
\hline
350&  1.014& 0.360&  0.461 \\
\hline
390&  0.994& 0.353&  0.452 \\
\hline
400&  0.982& 0.348&  0.447 \\
\hline
500&  0.900& 0.319&  0.409 \\
\hline
536&  0.864& 0.307&  0.393 \\
\hline
540&  0.859& 0.306&  0.391 \\
\hline
600&  0.792& 0.281&  0.360 \\
\hline
700&  0.651& 0.231&  0.296 \\
\hline
800&  0.482& 0.171&  0.219 \\
\hline
900&  0.343& 0.122&  0.156 \\
\hline
\end{tabular}

$^{1)}$ $\om_1=0,355$; $\om_2=0,455$.
\end{table}

\begin{figure}[h]
\centerline{
\epsfbox{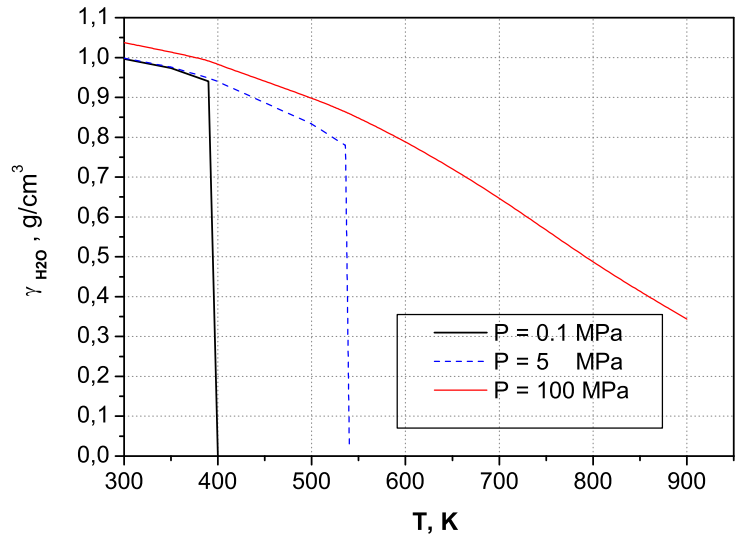} }
{\caption{\la{fig2}
Dependence of the water density $\ga_{H_2O}$  on the temperature
$T$ for different pressures $P$.     }}
\vspace*{0.5cm}
\end{figure}

\begin{figure}[h]
\centerline{
\epsfbox{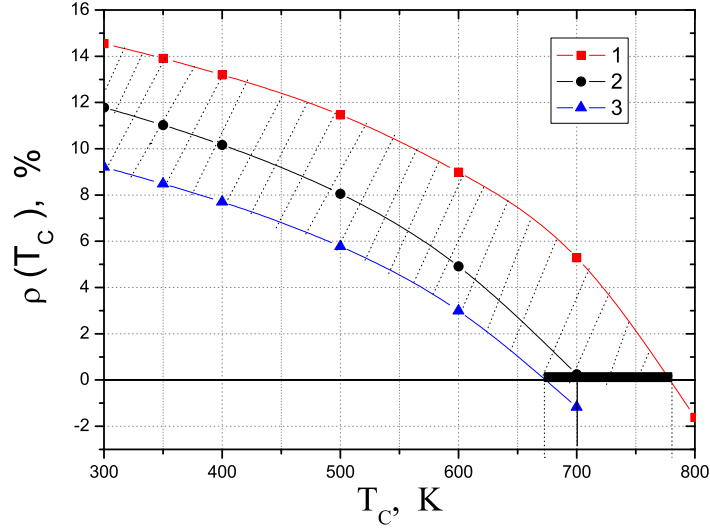} }
{\caption{\la{fig3}
{\em Power effect} \cite{35a,35b}. Dependence of the reactivity $\rho$
(in \%) of the fresh core of the Oklo reactor on the temperature $T_C$ at a
pressure of $P_C=100$ MPa for three different initial compositions of
the active core and different proportions of water: \\
{\em 1} -- 49.4 vol.\% U in ore, $\om_{H_2O}^0=0.455$;
{\em 2} -- 49.4 vol.\% U in ore, $\om_{H_2O}^0=0.405$;
{\em 3} -- 38.4 vol.\% U in ore, $\om_{H_2O}^0=0.355$.
The calculations were done using code MCU-REA. }}
\end{figure}

To determine the spread of results
depending on the uncertainty of the initial composition of the core the
calculations were carried out over a wide range of the content of
uranium ($Y_{U,i}(0)=39.4-59.6\%$ by weight) and of water
($\om_{H_2O}^0=0.355-0.455$) in the ore.  For reliability the
calculations for the bare reactor were carried out with two codes:
MCU-REA and MCNP4C.  The results were additionally controlled by the
single-group formula \ur{eq11} with the parameters shown in
Fig.\ref{fig4}. The calculations of the $T_C$ dependence
of  $K_{\rm eff}$
for variants {\em 2} and {\em 3} are similar (Fig.\ref{fig3}).  For
variant {\em 1} with lower uranium content ($Y_{U 1}(0)=38.4\%$ by
weight) the curve $K_{\rm eff}(T_C)$ lies visibly lower.  For a water
content of  $\om_{H_2O}^0=0.455$ the curve for variant {\em 2} lies
significantly higher. As a result the core temperature at which the
reactor became critical was $T_C=725$K with a spread of $\pm55$K.
Taking account of the fuel burn-up and of slogging of the reactor a
loss of reactivity takes place. This leads to a drop of $T_C$.  The
calculations of burn-up are continued at present.

\begin{figure}[h]
\centerline{
\epsfbox{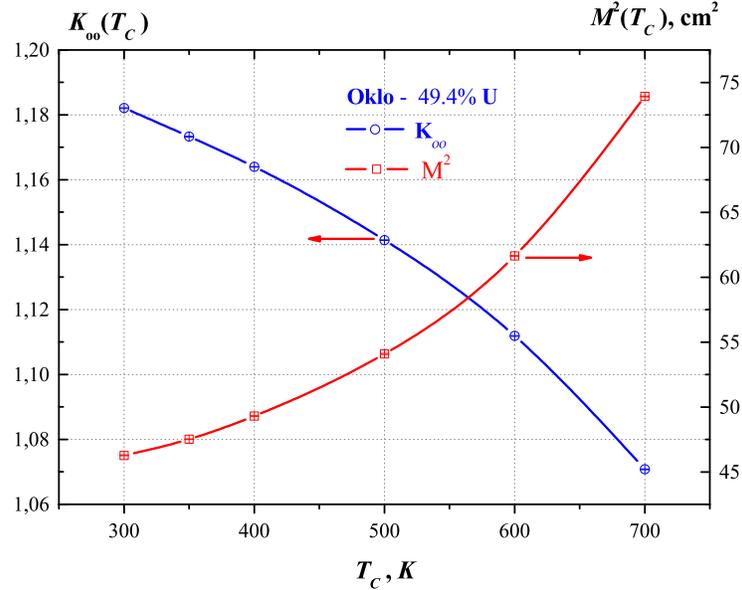} }
{\caption{\la{fig4}
$T_C$
dependence of $K_\infty(T_C)$ and $M^2(T_C)$ for a core containing
49.4 volume \% of uranium in the present-day dry ore and a water
content of $\om^0_{H_2O}=0.355$ at $T_C=300$K and $P_C=0.1$ MPa.
Calculations using code MCU-REA. \cite{35a, 35b} }}
\end{figure}

The
reactor could have worked also in a pulsating mode: when the
temperature exceeded 710~K, then the unbound water was boiled away and
the reactor stopped on account of the void effect. Then the water
returned and it started to work again \cite{27,36,37a,37b}.  However a
detailed analysis of the pulsating mode of operation of the reactor is
outside the scope of the present paper.

\subsubsection{Computational problems in calculations of large
reactors}

When making Monte Carlo calculations of the neutron flux in
large reactors, one encounters certain difficulties. In such reactors
many generations are produced before a neutron that was created in the
center of the reactor reaches its boundary. This time depends on the
relation between the migration length and the size of the reactor
\cite{38}.  At $T_C=300$K these values are for core $i=2$ equal to
$M=7$ cm and $R=8.1$ m. 230 generations are needed before a centrally
produced neutron reaches the boundary, detects the boundary condition
and returns to the center. In order to reproduce the spatial
distribution of the neutron flux with sufficient accuracy one must
calculate tens of such journeys.  Analyzing the solution of the time
dependent diffusion equation one finds that over 6000 cycles are needed
to get the fundamental harmonic with an accuracy of a few per cent.
Experience with such calculations shows that one needs
$(5-10)\cdot10^3$ histories per cycle in order to keep an acceptable
statistical accuracy. Thus we needed $(4-6)\cdot10^7$ neutron
trajectories for our calculations. For several hundred calculations we
have explored an order of 10$^{10}$ trajectories taking up several
months of continuous work of a modern PC cluster. In spite of such a
large volume of calculations we could not find the volume nonuniformity
coefficient $K_V$ of the neutron flux with good accuracy. To do these
calculations we had to divide the core into tens of volume elements
which led to a reduction of the statistical accuracy in each of them.
As a result the value of $K_V$ in formula \ur{eq12a} was reproducible
with an accuracy not better than 10\%.  This is obviously insufficient
to carry out the calculation of the burn-up that depends on the
magnitude of the absolute flux in different parts of the core.  One
must admit that the Monte Carlo method is not suitable for the
calculation of large reactors and one must resort to different
approaches.

The reactor neutron spectrum below  0.625 eV is needed in order to
average the cross sections of strong absorbers (e.g. $\Sm$).  The
spectrum for three compositions of the fresh core without reflector,
calculated with code MCNP4C for $T_C=300$K, is shown in Fig.\ref{fig5}.
One can see small peaks which correspond to excitations of rotational
and vibrational levels of H$_2$O. For comparison we also show in Fig.
\ref{fig5} the Maxwell neutron spectrum that was used by all previous
authors
to average the $\Sm$ cross section \cite{6a, 6b, 6c,7,8a,8b,9}. The
spectra are significantly different. The Maxwell spectrum has a much
higher peak but is exponentially small above 0.3 eV where the reactor
spectrum is a Fermi distribution. In our calculations we have used the
Nelkin model of water which automatically takes account of the chemical
bond of hydrogen nuclei. Calculations at other values of $T_C$ yield
similar results.

\begin{figure}[h]
\centerline{
\epsfbox{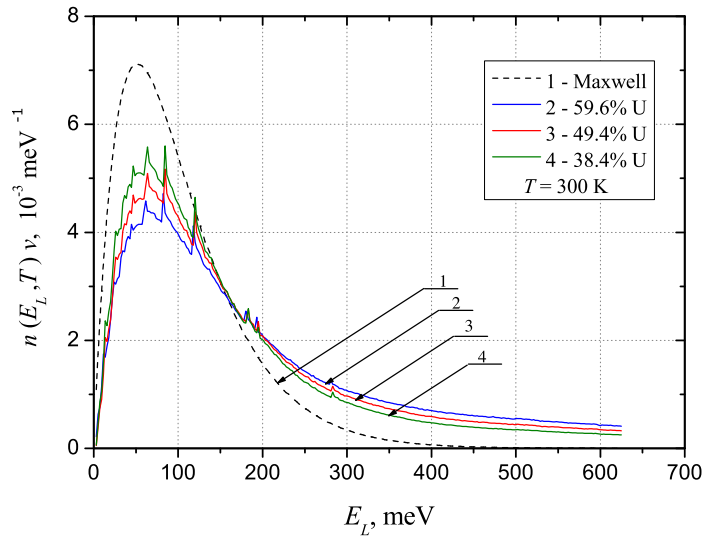} }
{\caption{\la{fig5}
Neutron spectrum in the bare fresh core RZ{\em 2}
at different initial uranium concentrations (water content
$\om_{H_2O}^0=0.355$).  Calculations using code MCNP4C
\cite{35a,35b}.}} \end{figure}

\section{Variation of fundamental constants}

The Oklo reactor is an instrument that is
sensitive to the neutron cross sections in the
distant past. Comparing them with current values one can
estimate how constant they, and hence also the fundamental
constants, are in time \cite{6a,6b,6c,7}.

\subsection{Early approaches}

In 1935 E. Miln posed
the question: how do we know that the fundamental constants are
actually constant in time \cite{39}. He thought that the answer could
be found only by experiment. A little later D. Dirac proposed  that
originally all constants were of one order of magnitude but that the
gravitational constant dropped at a rate of
$\dot{G}/G\sim -t_0^{-1}$ during the lifetime $t_0$ of the universe
\cite{40a,40b}.  In 1967 G. Gamov suggested that, on the contrary, the
electromagnetic constant is increasing: $\ah\sim t_0$ \cite{41}.  Both
hypothesis were wrong since they contradicted geological and
paleobotanical data from the early history of the Earth. Without
entering into a detailed discussion of these and many other later
publications on this subject one must admit that there is a problem of
the experimental limit on the rate of change of the fundamental
constants (see the early review by F.  Dyson \cite{42}).

The authors of Refs. \cite{6a,6b,6c, 7} noticed that the sensitivity
to variations of the nuclear potential increases by several orders of
magnitude if one considers neutron capture. Owing to the sharp
resonances of the absorption cross section the nucleus is a finely
tuned neutron receiver. A resonance shifts on the energy scale with
changes of the nuclear potential similarly as the frequency of an
ordinary radio receiver shifts when the parameters of the resonance
circuit are changed (Fig.\ref{fig6}) \cite{43}. Qualitatively one can
understand the absence of a significant shift of the near-threshold
resonances on the grounds that all strong absorbers are highly burnt up
in the Oklo reactor and weak absorbers are burnt up weakly
(Fig.\ref{fig7}) \cite{44,7}.  Holes in the distributions are seen for
strong absorbers:  $\mbox{}^{149}_{\:\:62}$Sm,
$\mbox{}^{151}_{\:\:63}$Eu, $\mbox{}^{155}_{\:\:64}$Gd,
$\mbox{}^{157}_{\:\:64}$Gd. The depth of burn-up, calculated using
the present absorption values, are in satisfactory agreement with
experiment, particularly if one remembers that the neutron spectrum
over which one must average the cross section is not known very well.
Thus in the 1.8 billion years since the work of the Oklo reactor, the
resonances (or, in other words, the levels of the compound nuclei) have
shifted by less than $\De E_r\si\Ga\ga/2$ ($\Ga_\ga=0.1$ eV).
Therefore the average rate of the shift did not exceed $3\cdot10^{-11}$
eV/year.  This value is at least three orders of magnitude less than
the experimental limit on the rate of change of the transition energy
in the decay of $\mbox{}^{187}$Re \cite{42}.

\bigskip
\bigskip
\begin{figure}[h]
\centerline{
\epsfxsize=9.5cm
\epsfysize=8cm
\epsfbox{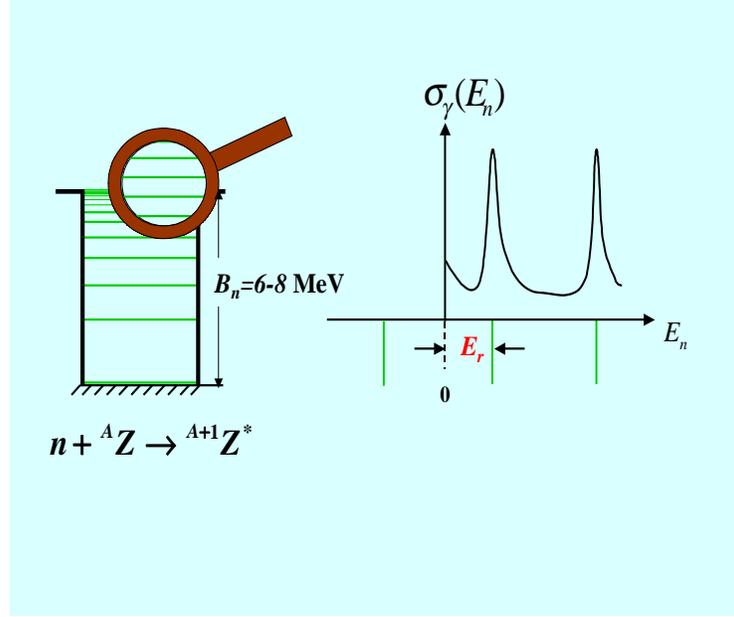} }
\caption{\la{fig6}
Strong absorber as a sensitive detector of a variation of $E_r$
\cite{43}.  Shown on the left is the energy level density of the
compound nucleus $n+^AZ\to^{A+1}Z^*$. Shown on the right are resonances
in the cross section of the reaction . The capture cross section
behaves like $\sig_\ga\simeq (\Ga_\ga/E_r)^2$, where $E_r$ is the
distance from the resonance and $\Ga_\ga$ is its width.  Neutron
capture is strongly affected by a shift of $\De E_r$.} \end{figure}

\begin{figure}[h]
\vspace*{0.5cm}
\centerline{
\epsfbox{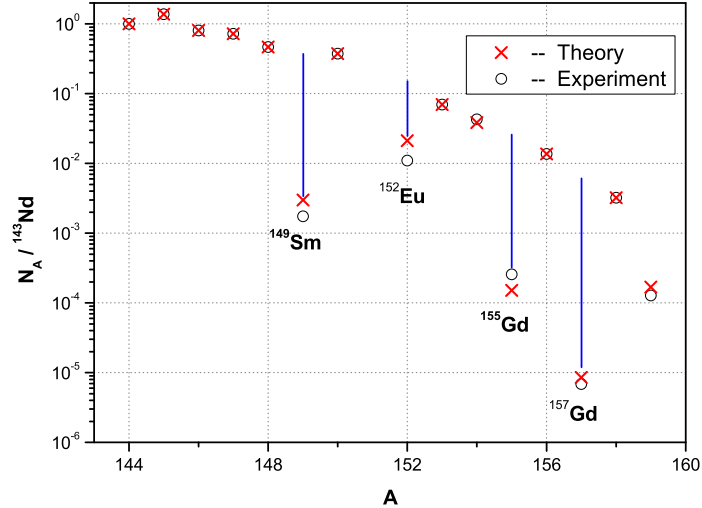} }
{\caption {\la{fig7}
Comparison of the calculated (crosses) and measured (
circles) concentrations of fission fragments $N_A$ relative to the
$\mbox{}^{143}$Nd content for one of the samples of the Oklo reactor
\cite{44, 7}.}}
\end{figure}

At
present there are no theoretical calculations giving a reliable
connection between the positions of all resonances with parameters of
the nuclear potential.  But already the preliminary qualitative
estimates allow one to reduce the limits on the rates of change of the
coupling constants of the strong and electromagnetic interactions
$\dot{\overline{\alpha}}/\alpha$ and $\ad$. We confirm the absence of a
power or logarithmic dependence on the lifetime of the universe.  It is
desirable to have a more detailed calculation of the influence of
variations of the fundamental constants on the parameters of the
neutron resonances.

\subsection{Basic formulae}

\subsubsection{Averaging the Breit-Wigner formula}

When a slow neutron is captured by
a nucleus of isotope $\Sm$, then a nuclear reaction takes place with
formation of an excited intermediate compound nucleus and subsequent
emission of $m$ $\ga$--quanta:
\beq
n+\Sm\to\mbox{}^{150}_{\:\:62}{\rm Sm}^*\to\mbox{}^{150}_{\:\:62}{\rm
Sm}+m\cdot \ga\, .
\eeq
Near a strong $S$ resonance one can neglect the
effect of the other resonances and describe the cross section with the
Breit-Wigner formula
\beq
\sig_{\ga,Sm}(E_C)=g_0\frac{\pi\planck^2}{2m_n
E_C}\frac{\Ga_n(E_C)\cdot\Ga_\ga}{(E_C-E_r)^2+\Ga_{tot}^2/4},
\la{eq15}
\eeq
where $g_0=(2J+1)\bigl/(2S+1)(2I+1)$ is the statistical factor, $S=1/2$
is the electron spin, $I$ is the nuclear spin and $J$ is the spin of
the compound nucleus.  The full width is $\Ga_{tot}=\Ga_n(E)+\Ga_\ga$,
where $\Ga_n(E)$ and $\Ga_\ga$ are the neutron and $\ga$  width,
respectively.  The neutron width is given by \cite{45}
\beq
\Ga_n(E_C)=\Ga^0_n\sqrt{\frac{E_C}{E_0}}; \quad E_0=1 \, {\rm eV}.
\eeq
The
parameters of the lowest resonances of a number of absorbers is given
in Table \ref{tab9}.

\begin{table}[h]
\caption{\la{tab9}
Lowest resonance parameters of strong absorbers \cite{45}}

\begin{tabular}{|l|c|c|c|} \cline{2-4}
\multicolumn{1}{c|}{}& $^{149}_{62}$Sm& $^{155}_{64}$Gd&
$^{157}_{64}$Gd \\  \hline
Resonance energy ($E_r$), meV& 97.3& 26.8& 31.4 \\ \hline
Neutron width at 1 eV ($\Ga_n^0$),meV&
1.71(3)& 0.635(10)& 2.66(5) \\ \hline
Gamma width ($\Ga_\ga$), meV&
60.5(6)& 108(1)& 106(1) \\ \hline
g -- factor & 9/16 & 5/8& 5/8 \\ \hline
Doppler width ($\Delta_D$), meV at $T=700$ K&
12.5& 6.4& 6.9 \\
\hline
\end{tabular}
\end{table}

In formula \ur{eq15} the neutron energy is given in the c.m. frame:
$E_c=\frac{1}{2}m_n\mid\vec V_L-\vec V_k\mid^2$. It depends on the
velocities of the nucleus $\vec V_k$ and the neutron $\vec V_L$ in the
lab frame and on the reduced mass $m_n$.  The reaction rate
$N_k\sig_{\ga,k}(E_C)\cdot V_C$ with cross
section \ur{eq15} and for an absorber of nuclear density $N_k$ must be
averaged over the nuclear spectra $f_k(E_k)$ and the neutron spectrum
$n(E_L)$ (all spectra are normalized to one).  The inverse nuclear
burn-up time in an arbitrary point of the core is given by
\beq
\lambda_{\ga,k}(T)=N_k\int\!\! d\vec p_k d\vec p_L
f_k(E_k)n(E_L)\sig_{\ga,k}(E_C)V_C\, .
\eeq
At high temperatures the
gas approximation is valid for heavy nuclei of the absorber.  Changing
to integration over the c.m. energy $E_C$ and the neutron energy $E_L$
and assuming a Maxwell nuclear spectrum, we get
\beq
\lambda_{\ga,k}(T)=N_k\int n(E_L)\sig_{\ga,k}(E_C)V_C F(E_C\to E_L)d E_L
d E_C\, ,
\la{eq18}
\eeq
$F(E_C\to E_L)$ is the
transformation function from the c.m. to the lab system (for details
see Ref. \cite{46}):
\beq
F(E_C\to E_L)=\frac{(A+1)}{2\sqrt{\pi A T  E_L}} \left\{
\exp\left[  -\frac{A}{T}\left(\sqrt{(1+\frac{1}{A})\cdot
E_C}-\sqrt{E_L}\right)^2\right]-
\exp\left[
-\frac{A}{T}\left(\sqrt{(1+\frac{1}{A})\cdot
E_C}+\sqrt{E_L}\right)^2\right]
\right\}
\la{eq19}
\eeq
and $A=M_A/m_n$  is the mass of nucleus $A$ in
units of the neutron mass.

Close to a resonance we can neglect the
 second term in Eq.\ur{eq19} and evaluate the first term in integral
\ur{eq18} by the saddle-point method.  As a result the integral
\ur{eq18} takes on the following form in the vicinity of a resonance:
\beq
\lambda_{\ga,k}(T)=N_k\frac{\pi}{2}\left(1+\frac{1}{A}\right)\int\!\!
dE_C \sigma_{\ga,k}(E_C)V_C \int d E_L n(E_L) \Ga\left[E_L-
\left(1+\frac{1}{A}\right)
E_C
\right]\, ,
\la{eq20}
\eeq
where the Gaussian
\beq
\Ga\left[E_L-
\left(1+\frac{1}{A}\right)
E_C\right]
=
\frac{1}{\sqrt{\pi}\De_D}
\exp\left\{-\frac{
\left[E_L-\left(1+\frac{1}{A}\right)E_C
\right]^2}
{\Delta_D^2}
\right\}
\la{eq21}
\eeq
is normalized to one and the Doppler width is equal to
\beq
\De_D=\left[\frac{4E_LT}{A}\right]^{1/2}
=\left[\frac{4E_CT}{A+1}\right]^{1/2}
\eeq
The values of the Doppler widths for $T=700$K are shown in Table
\ref{tab9}.  Since all $\De_D\ll\Ga_\ga$, function \ur{eq21} can be
replaced by $\delta\left(\mid E_L-AE_C/(A+1)\mid\right)$ and integral
\ur{eq20} becomes \beq
\lambda_{\ga,k}=N_k\frac{\pi}{2}\left(1+\frac{1}{A}\right)^2
\int
\sigma_{\ga,k}(E_C) V_C n \left[\left(
1+\frac{1}{A}\right)E_C\right]dE_C\, .
\la{eq23}
\eeq
The correction $2/A$ is of magnitude 1\%. If the neutron spectrum is
Maxwellian in the c.m. frame, then it is also Maxwellian (with
reduced neutron mass) when the nuclear motion is taken into account.
It can be shown that Eq.\ur{eq23} is valid at all energies if the
distribution of nuclei and neutrons is Maxwellian \cite{47}.

Therefore it is not surprising that the authors of Ref. \cite{9} did
 not notice any deviations from formula \ur{eq23}
in their numerical
 calculation that took account of the thermal motion of the target
nuclei (Doppler effect).  However, the situation is different if the
neutron spectrum is not Maxwellian. In this case one must use formula
\ur{eq20} instead of the simple formula \ur{eq23}.

To average the capture cross
section of samarium one normalizes the cross section, integrated over
the neutron flux spectrum $(n(E,T)v)$, traditionally not by the
integrated flux but by the product of the velocity $v_n^0=2200$ m/s and
the integrated neutron density $(n(E))$ \cite{47,8b,9}:
\beq
\hat\sig_{\ga,k}(T)=\frac{\int\sig_{\ga,k}(E_L)n(E_L)\cdot v_L d
E_L}{v_n^0\int n(E_L)dE_L}\, .
\la{eq24}
\eeq
If the cross section $\sig_{\ga,k}(E_L)$ has a $1/v_L$ behaviour, then
the integral of $\hat\sig_{\ga,k}(T)$ is constant.  From formula
\ur{eq24} one has
\beq
\hat\sig_{\ga,k}(T)=\sqrt{\frac{4T}{\pi T_0}}\sig_{\ga,k}(T)\, ,
\la{eq25}
\eeq
where $T_0=300$ K $=25.9$ meV. Useful is also the relation \cite{9}
\beq
\sig\Phi=\hat\sig\hat\Phi, \quad {\rm where} \quad \hat\Phi=
\sqrt{\frac{\pi}{2}\frac{T_0}{T}}\Phi\, .
\eeq

We have evaluated the cross section $\hat\sig_{\ga,Sm}(T)$ of $\Sm$
without recourse to any approximations. For
$\lambda_{\ga,Sm}(T)/(N_{ Sm}v^0_n)$ we have
\beq
\hat\sig_{\ga,Sm}(T)=\frac{\sqrt{\pi}\int\!\! dE_kdE_L f_{Sm}(E_k)
\sig_{\ga,Sm}(E_C)\cdot V_C n(E_L)}{v_n^0\int\!\!dE_L n(E_L)}\, .
\la{eq27}
\eeq
For the calculations we used the computer package MATHEMATICA
\cite{48}.  In Fig.\ref{fig8} we show the values of
$\hat\sig_{\ga,Sm}(T,\De E_r)$ at six temperatures $T=300-1000$K for a
shift of the resonance position $\De E_r=\pm0.2$ eV.  The curves have a
maximum at negative shifts of the resonance; the maximum of the curves
is higher at lower temperature $T$.
At the point $\De E_r=0$ and at
$T=293$K the cross section calculated as
the contribution of the closest resonance
is equal to $\sig_{\ga,Sm}(293\,\text{K})=39.2$ kb.
The contribution of higher positive resonances is
$\sig_{\ga,Sm}^+(293\,\text{K})=0.6$ kb and negative ones
is $\sig_{\ga,Sm}^-(293\,\text{K})=0.3$ kb \cite{45}.
The total cross section (as measured on a neutron beam)
is  $\sig_{\ga,Sm}^{tot}(293\,\text{K})=40.1$ kb.
At small energy shifts $\De E_r$
$\sig_{\ga,Sm}^+$ and $\sig_{\ga,Sm}^-$ practically do not
change. In Refs.\cite{8a,8b,9} the total cross section
$40.1$ kb has been used instead of the single resonance
one $39.2$ kb.Therefore the curves in Refs. \cite{8a,8b,9} are
higher by 40.1 kb/39.2 kb, i.e. by  2.5\%.

\begin{figure}[h]
\centerline{
\epsfbox{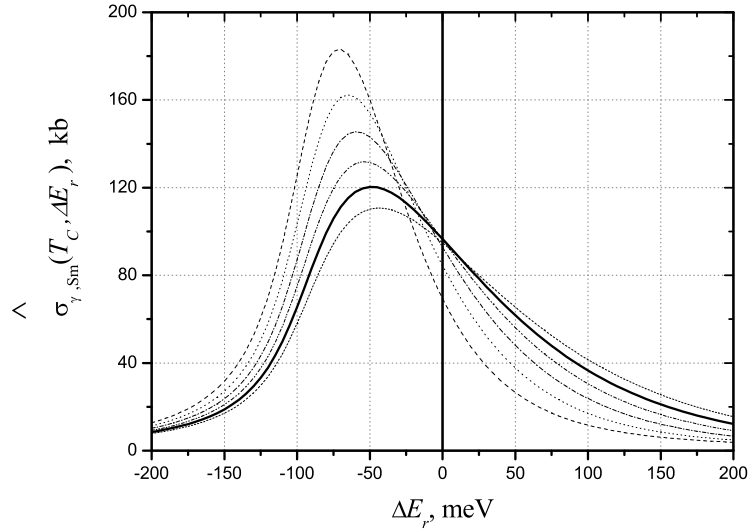} }
{\caption{\la{fig8}
Dependence of the cross section $\hat\sig_{\ga Sm}$, averaged
over a Maxwell neutron spectrum, on the resonance shift $\De E_r$  and
on the temperature: $T=(300-800)$K \eq{eq27}. The curves are for
fixed temperatures with intervals of 100 K. The upper left curve is for
$T=300$K. }}
\end{figure}

\subsubsection{Taking account of the reactor spectrum}

\la{2.2.2}
In Fig.\ref{Fig9} we
show the results of calculating $\hat\sig_{\ga,Sm}(T_C,\De E_r)$  with
the Maxwell spectrum replaced by the reactor spectrum
$n_R(E_L,T_C)$.  The central curve {\em 2} is the result of the
calculation using code MCNP4C for the fresh core with
$Y_{U2}(0)=49.4\%$U in the dry ore and with $\om^0_{H_2O}=0.405$ at
$T=725$K.  For comparison we also show the cross section averaged
over the Maxwell spectrum at $T=725$K for the same composition of
the core (curve {\em 4}).  Curves {\em 4} and {\em 2} are significantly
different, especially at negative $\De E_r$:  curve {\em 2} lies
distinctly lower.  The maximum of curve {\em 2} is 1.5 times lower than
the maximum of curve {\em 4}.  At lower temperatures this difference is
even greater. Thus we conclude that we have proved a significant
{\em effect of the reactor spectrum on the cross section of} $\Sm$.

\begin{figure}[h]
\centerline{
\epsfbox{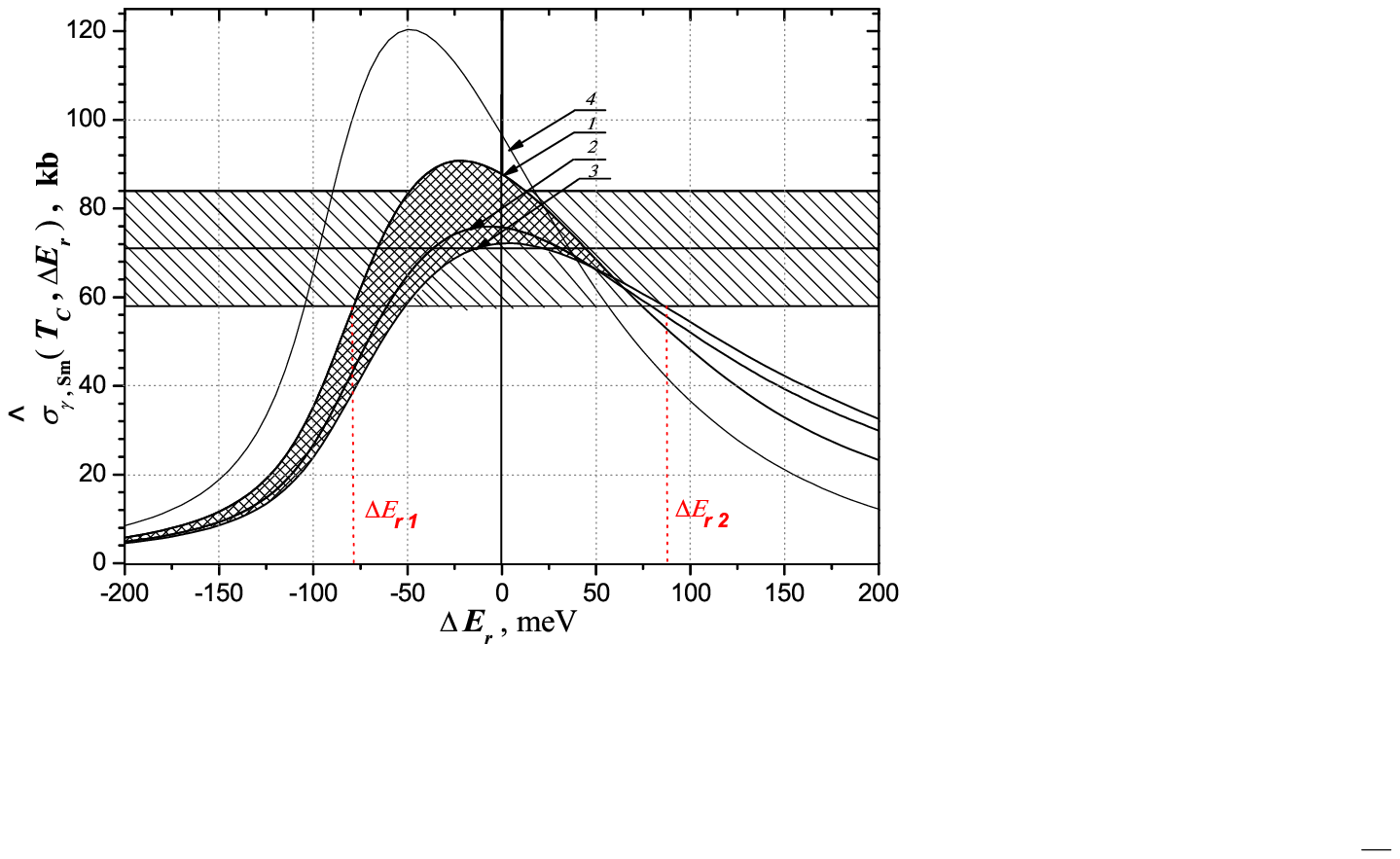} }
{\caption{\la{fig9}
Dependence of the thermal neutron capture cross section of
isotope $\Sm$ on the core temperature $T_C$ and on the resonance shift
$\De E_r$: $\hat\sig_{\ga, Sm}(T, \De E_r)$ \cite{35a,35b}.  The
curves are for cross sections, averaged over the reactor spectrum of
the fresh core at three different initial states:\, {\em 1} -- $Y_{U
2}(0)=38.4$ vol.\%U, $\om_{H_2O}=0.355$, $T_C=670$K; \, {\em 2} --
$Y_{U 2}(0)=49.4$ vol.\%U, $\om_{H_2O}=0.405$, $T_C=725$K; \, {\em 3}
-- $Y_{U 2}(0)=49.4$ vol.\%U, $\om_{H_2O}=0.455$, $T_C=780$K, $P_C=100$
MPa.  \, For comparison we show the cross section averaged over the
Maxwell spectrum (curve {\em 4}) for initial composition $Y_{U
2}(0)=38.4$ vol.\%U, $\om_{H_2O}=0.405$, $T_C=725$K.  Shown is the
error corridor of measured values $\hat\sig_{\ga, Sm}^{\rm Exp}$
\cite{8a,8b}.}}
\end{figure}

\noindent
In order to determine the dependence of $\hat\sig_{\ga,Sm}(T_C,\De E_r)$
on the uncertainty in the
initial active core composition, we have calculated the values for the
two outermost curves of Fig.\ref{fig9}. Curve {\em 3} of this figure
corresponds to an initial content of $Y_{U1}(0)=38.4\%$U in the ore,
$\om^0_{H_2O}=0.355$  and $T_C=670$K; curve {\em 1} corresponds to an
initial content of 49.4\%, $\om^0_{H_2O}=0.455$ at $T=780$K.  Since
the numerical constants are known only for values of $T_C$ which are
multiples of 100, we have done the calculations for
$T_C=(600,700,800)$K and interpolated to intermediate temperatures.
The broadening of curve {\em 2} on account of the scatter of
temperatures is small. Experimental data of
$\hat\sig_{\ga,Sm}^{\rm Exp}(T)$  for core {\em 2} are presented
in Ref. \cite{8a} (Table \ref{tab10}) (see also Refs. \cite{52}). The
labeling of sample SC36-1418 indicates that the sample was taken from
bore-hole SC36 at a depth of 14 m 18 cm. The mean value
\la{xxx}
$\ov{\hat\sig}_{\ga,Sm}^{\rm Exp}=(73.2\pm 9.4)$ kb is shown in
Fig.\ref{fig9}.  Curve {\em 1} ($T_C=670$K) crosses the lower limit of
$\ov{\hat\sig}_{\ga,Sm}^{\rm Exp}=64$ kb to the left  of point
$\De E_r^{(1)}=-73$ meV, and curve {\em 3} ($T_C=780$K) to the
right at $\De E_r^{(2)}=+62$ meV.  The possible shift of the
resonance is therefore given by these limits:
\beq
-73 \, {\rm meV} \,\leq \De E_r\leq 62 \, {\rm meV}.
\la{eq28}
\eeq

\begin{table}[h]
\caption{\la{tab10}
Experimental values of $\hat{\sig}_{\rm Sm}$ for 15
samples from Oklo reactor core \RZ}

\begin{tabular}{|c|l|c|c|} \hline
&&  $\hat{\sig}_{\rm Sm}$, kb& \\
\hline
1&  KN50-3548&     93&   \cite{44} \\
\hline
2&  SC36-1440&     73&   \cite{49} \\
\hline
3&  SC36-1410/3&   73&   \cite{49} \\
\hline
4&  SC36-1413/3&   83&   \cite{49} \\
\hline
5&  SC36-1418&     64&   \cite{49} \\
\hline
6&  SC39-1383&     66&   \cite{20b,49}  \\
\hline
7&  SC39-1385&     69&   \cite{20b,49}  \\
\hline
8&  SC39-1389&     64&   \cite{20b,49}  \\
\hline
9&  SC39-1390&     82&   \cite{20b,49}  \\
\hline
10& SC39-1391&     82&   \cite{20b,49}  \\
\hline
11& SC39-1393&     68&   \cite{20b,49}  \\
\hline
12& SC35bis-2126&  57&   \cite{20b,49}  \\
\hline
13& SC35bis-2130&  81&   \cite{20b,49}  \\
\hline
14& SC35bis-2134&  71&   \cite{20b,49}  \\
\hline
15& SC52-1472&     72&   \cite{25}      \\
\hline
  & $\ov{\hat{\sig}}_{\rm Sm}\pm\De\ov{\hat{\sig}}_{\rm Sm}$, kb &
(73.2$\pm$9.4)& \\
\hline
\end{tabular}
\end{table}

\subsubsection{Connection between $\De E_r$ and $\ad$}

The shift $\De E_r$ must be related to
a  variation of the fundamental constants, for instance to a shift of
the electromagnetic constant $\al=1/137.036$. This has been
done by Damour and Dyson \cite{8a,8b}. The change of the Coulomb energy
contribution $\De H_C$ to the energy of the level in the nuclear
potential $\De E_C$, that results from a change of $\al$, is given by
\beq
\frac{\partial}{\partial \alpha}\De E_C
=\frac{\partial}{\partial \alpha}\langle \Delta H_C\rangle\quad.
\eeq
In first perturbative approximation the dominant contribution to
$\De E_C$ is given by the isotopic effect (\cite{51}, p.568)
\beq
\De
E_C=E_R=\frac{2\pi}{3}\mid\psi_e(0)\mid^2 Z\cdot e^2
<R^2>\, .
\eeq
Here $\psi_e(0)$ is the wave function of the $s$ wave electrons in the
nucleons and $\langle R^2\rangle=(Z\cdot e)^{-1}\int\rho R^2 dV$, where
$\rho(R)$ describes the proton charge distribution in the nucleus.
Damour and Dyson have estimated the value of $\langle R^2\rangle$ for
the excited nucleus $\mbox{}^{150}_{\:\:62}$Sm$^*$ from the
neighbouring isotopes.  They found
\beq
{\cal M}=\De E_R=-(1.1\pm0.1) \, {\rm
MeV}\, .
\eeq
Combining this
value with the shift of the resonance $\De E_r$ in formula \ur{eq28},
we get
for $\be\De E_r/{\cal M}$
\beq
-5.6\cdot10^{-8}<\at<6.6\cdot10^{-8}\, .
\eeq
Because of the negative value of ${\cal M}$ the limits on $\at$
change their places.  For the past time ($-T_0$) the product
($-T_0{\cal M}$) is positive and hence the limits on $\ad\equiv\De
E_r(-T_0{\cal M})$ are restored to their previous places.  Note that
traditionally $\at$ is defined by
$\de\al=(\al_{Oklo}-\al_{now})\bigl/\al$.  This shift of $\al$ lies in
a narrower range than in Ref. \cite{8a,8b}.  Assuming a linear change of
the e.m.  constant during the time $T_0$, we get the following limit on
the relative rate of change:
\beq
-3.7\cdot10^{-17}\, {\rm year}^{-1}\,
<\ad<3.1\cdot10^{-17}\, {\rm year}^{-1}\, .
\la{eq33}
\eeq
Thus, within the limits given by Eq. \ur{eq33}, the e.m.  constant changes
for the fresh reactor Oklo core with zero speed, i.e. it remains
constant.

\subsection{Review of previous work}

{\em The work of Shlyakhter (1976, 1983) \cite{6a,6b,6c}}.
The authors of Refs. \cite{6a,6b,6c, 7} were the first to point out the
possibility of using the data of the natural nuclear reactor Oklo to
find the most precise limits on the rate of change of the fundamental
constants. The most convenient data are those of the strong absorbers,
e.g. of $\Sm$. For this isotope Shlyakhter calculated at $T=300$K the
dependence of the change of the cross section on the resonance by an
amount of $\De E_r$: $\sig_{\ga,Sm}(T_C,\De E_r)$ (Fig.\ref{fig10})
\cite{6b}.

\begin{figure}[h]
\centerline{
\epsfbox{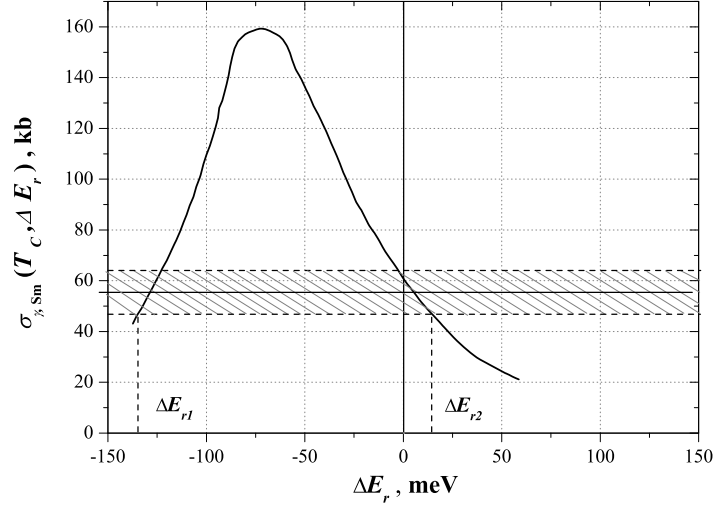} }
{\caption{\la{fig10}
Change of the thermal ($T=25$meV$=273$K) capture cross
section of $\Sm$ for a uniform shift of all resonances by $\De E_r$
\cite{6b}.}}
\end{figure}

\noindent
He
compared this curve with the experimental data available at the time:
$\ov\sig_{\ga,Sm}^{\rm Exp}=(55\pm8)$ kb.
A possible shift of the first resonance within two standard deviations
(95\% confidence level) was found to be
\beq
\de E_{r 2}^{\rm Exp}\leq 20 \, {\rm meV}\, .
\eeq
(In going from $\ov\sig_{\ga,Sm}(T,\De E_r)$ to $\hat\sig_{\ga,Sm}(T,\De
E_r)$, all values must be multiplied by 1.18 according to Eq.
\ur{eq25} and $\ov{\hat\sig}_{\ga,Sm}^{\rm Exp}=(65\pm9.5)$ kb, but
this does not affect the value of $\de E_{r 2}^{\rm Exp}$). To estimate
${\cal M}$, Shlyakhter used data on the compressibility of the nucleus;
he found ${\cal M}=-2$ MeV \cite{6b}.  With a linear dependence of
the change of $\al$ with time for $T_0=2\cdot10^9$ years the limit on
the rate of change of $\al$ is \beq \ad\le 0.5\cdot10^{-17} \, {\rm
year}^{-1}\, .
\eeq
Using the present, more accurate value of
${\cal M}\si -1.1$ MeV, we get
\beq
\ad\le 1\cdot10^{-17} \, {\rm year}^{-1}\, .
\eeq
It should emphasized again that this limit was found for
only one temperature: $T=300$K.

{\em The work of Petrov (1977) \cite{7}}.
In this paper the resonance shift $\de E_r$ was estimated from
the shifts of the widths of a few
strong absorbers. The resonances of strong absorbers lie close to a
zero energy of the neutron, and the resonance energy is of the capture
width: $\de E_r\sim \Ga_\ga\si 0.1$. The capture cross
section of these absorbers for thermal neutrons changes sharply when
the resonance is shifted by an amount of the order of $\Ga_\ga/2$. The
analysis of experimental data for $\Sm$ and $\mbox{}^{151}_{\:\:63}$Eu,
taking account of a threefold standard deviation and the uncertainly of
the core temperature, shows that the shift $\de E_r$  of the resonance
since the activity of the Oklo reactor does not exceed $\pm$0.05 eV
\cite{7}.  The results of the measurement of the concentration of
rare earth elements with respect to $\mbox{}^{143}_{\:\:60}$Nd (the
second branch of the mass distribution of the fission fragments) in one
of the Oklo samples are shown in Fig.\ref{fig7}.  A more conservative
estimate in Ref. \cite{7} is $\mid\ov{\de E_r}\mid\leq50$ meV, i.e.
2.5 times higher than Shlyakhter's estimate.  Using the modern value
${\cal M}\si-1.1$ MeV, we find for $\at\leq\ov{\de E_r}\bigl/\mid
{\cal M}\mid$
\beq
\at=\ov{\de E_r}\bigl/\mid{\cal M}\mid\leq4.5\cdot10^{-8}\, .
\eeq
This is almost 5 times greater than Shlyakhter's optimistic estimate.
For the rate of change $\ad$ we get
\beq
\ad\leq2.5\cdot10^{-17} \, {\rm year}^{-1} \, .
\eeq
This is less by a factor of 2 than the limit found 20 years later
by Damour and Dyson \cite{8a,8b}. The reason of this discrepancy is the
use of only one temperature ($T_C=300$K).  Although the dependence of
$\De E_r$ on $T_C$ was noted in Ref. \cite{7}, no calculations of the
effect of the temperature were carried out.

{\em The work of Damour and Dyson (DD) (1996) \cite{8a,8b}}.
The dependence $\De E_r(T_C)$  was analysed 20 years later in the paper
DD \cite{8a,8b}.  They have repeated the analysis of Shlyakhter and came
to the conclusion that it was correct. DD also updated Shlyakhter's
data in three directions:   \\
({\em i})  They employed a large amount of experimental data
(see Table \ref{tab10}). \\
({\it ii}) They have taken account of the
great uncertainty of the reactor temperature, $T_C=(450-1000)^0$C (Fig.
\ref{fig11}). As a result they made a conservative estimate of the mean
shift of the resonance
\beq -120 \, {\rm meV} \,\leq {\ov\De E_r}\leq
90 \, {\rm meV}.
\eeq
The range of the shift $\De E_{r1}-\De
E_{r2}=210$ meV is 1.5 times greater than the range of the shift in our
paper.  \\ ({\it iii}) They have calculated the value
${\cal M}=-(1.1\pm0.1)$ MeV, but used ${\cal M}=-1$ MeV.
For $\at$ DD
found
\beq -9.0\cdot10^{-8}\leq\at\leq12\cdot10^{-8}\, .
\eeq
This
leads to the following limits on the rate of change $\ad$:
\beq
-6.7\cdot10^{-17} \, {\rm year}^{-1} \leq\ad\leq5.0\cdot10^{-17}\,
{\rm year}^{-1} \, .
\eeq
Since $\Sm$ burns up 100 times faster than $\U$, therefore the only
$\Sm$ found in the stopped reactor is that which was produced
immediately before the end of the cycle. As a consequence DD emphasized
that one must know the detailed distribution of the nuclear reaction
products of the end of the cycle to make a detailed analysis.

\begin{figure}[h]
\centerline{
\epsfbox{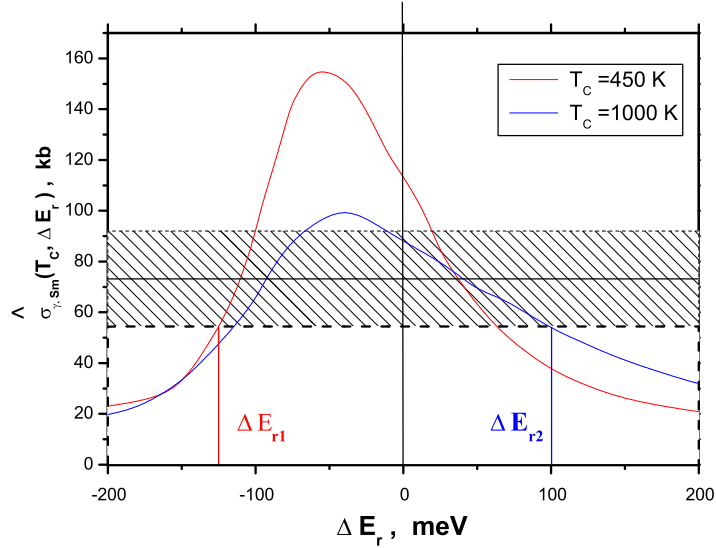} }
{\caption{\la{fig11}
Dependence of the thermal neutron capture cross section on the
resonance shift $\De E_r$ for the two temperatures of the
Maxwell neutron spectrum ({\em Damour $\&$ Dyson} \cite{8a,8b}).
The energies of the crossing of the lower limit of data are
$\De E_{r 1}^{M}=-120$ meV and $\De E_{r 2}^{M}=90$ meV.  }}
\end{figure}

{\em The work of Fujii et al.  (2000, 2002) \cite{9, 53}}.
The experimental data on the measurement of $\hat\sig_{\ga,k}$ of
strong absorbers were presented in the papers of Fujii et al. Of five
experimental points, four are from core {\em RZ10\ } (Table
\ref{tab11}) and one from another core, {\em RZ13\ }, and therefore we
have omitted it.  Core {\em RZ10\ } lies at a depth of about 150 m from
the surface of the quarry.  As we do not have any detailed data on the
size and composition of core {\em RZ10\ }, we shall assume them to be
similar to those of core \RZ. For $\Sm$  the mean value of the four
points of Table \ref{tab11} is
\beq
\ov{\hat\sig}_{\ga,Sm}^{\rm
Exp}=(90.7\pm8.2) \, {\rm kb}\, .
\la{eq42}
\eeq
This value is
noticeably greater than for core \RZ [(72.3$\pm$9.4) kb (see subsection
\ref{xxx}). The error bars of both values do not even touch and their
difference remains significant. For increased reliability the number of
measurements for core {\em RZ10\ } should be increased. Possibly this
core has finished its cycle at a lower temperature.

\begin{table}[h]
\caption{\la{tab11}
Experimental values of $\hat{\sig}^{Exp}_{\ga,k}$ (kb) for
strong absorbers, measured in the middle of the core {\em RZ10\ },
located at a depth of 150 m \cite{9} }

\begin{tabular}{|c|c|c|c|c|} \cline{3-5}
\multicolumn{2}{c|}{} & $^{149}_{62}$Sm& $^{155}_{64}$Gd&
$^{157}_{64}$Gd \\  \hline
 &  & $\hat{\sig}_{149}^{\rm Sm}$, kb& $\hat{\sig}_{155}^{\rm Gd}$, kb&
$\hat{\sig}_{157}^{\rm Gd}$, kb  \\    \hline
1& SF84-1469&  83.6&  30.9&   83.3   \\
2& SF84-1480&  96.5&  16.8&   8.0    \\
3& SF84-1485&  83.8&  17.8&  14.3    \\
4& SF84-1492&  99.0&  36.7&  73.7    \\ \hline
5& $\ov{\hat{\sig}}_k$&  90.7& 25.6& 44.8 \\
6& $\pm\De\sig_k$&  8.2&  9.8& 39.2    \\ \hline
\end{tabular}
\end{table}

In Fig.\ref{fig12} we show the dependence of $\hat\sig_{\ga,Sm}(T_C,\De
E_r)$ on the shift of the resonance for a Maxwell distribution \cite{9,
52}.  The authors have estimated (on the grounds of indirect
considerations) the uncertainty in $T_C=(180-400)^0{\rm C}=(453-673)$K.
They also show the experimental data of formula \ur{eq42}. The
intersection
of the limiting curves with the lower limit $\hat\sig_{\ga,Sm}^{\rm
Exp}=82.5$ kb yield the following possible shift of $\De E_r$:
\beq
-105 \, {\rm meV}< \De E_r<+20\, {\rm meV}\, .
\la{eq43}
\eeq
In Fig.\ref{fig11} we also show for comparison the curves
$\hat\sig_{\ga,Sm}(T_C,\De E_r)$ for the reactor
spectrum of the fresh core at $T_C=400^0$C, $Y_{U1}(0)=38.4\%$U in the
ore, $\om^0_{H_2O}=0.355$  and $P=100$ MPa.  The possible shift of
$\De E_r$ for the experimental data of formula \ur{eq42}
lie in a narrower interval than in the paper DD \cite{8a,8b}:
\beq
-120 \, {\rm meV}\leq\De E_r\leq20\, {\rm meV}\, .
\eeq
From  Eq.\ur{eq43} and using $\M=-1.1$ MeV we get
\beq
-1.8\cdot10^{-8}\leq\ov{\de\al/\al}\leq9.5\cdot10^{-8}\,
\eeq
and
\beq
-5.3\cdot10^{-17} \, {\rm year}^{-1} \leq
\dot{\ov{\de\al/\al}}\leq1.0\cdot10^{-17}\, {\rm year}^{-1} \, .
\eeq
Thus in this case too we do not find with certainty a
nonzero deviation of the
change  of $\al$.

\begin{figure}[h]
\centerline{
\epsfbox{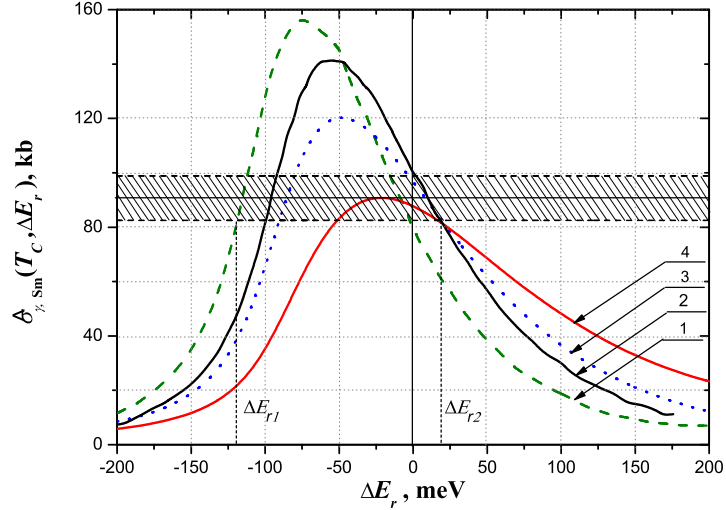} }
{\caption{\la{fig12}
Dependence $\hat\sig_{\ga, Sm}(T_C,\De E_r)$ for the
Maxwell distribution at $T_C=(450,570,670)$K ({\em curves 1--3})
$\bigl($Fujii et al.\cite{9,53}$\bigr)$.  The data correspond to
Table \ref{tab11}; $\hat\sig_{\ga Sm}^{\rm Exp}=(90.7\pm8.2)$ kb,
$\De E_{r 2}^{M}=20$ meV.  For comparison we show {\em curve 4}
for the reactor spectrum at $T_C=670$K; $Y_{U 1}(0)=38.4$ vol.\%U;
5$\om_{H_2O}^0=0.355$; $P_C=100$ MPa. }}
\end{figure}

{\em The work of Lamoreaux and Torgerson (LT) (2004) \cite{53,54}}.
These authors noted (two years after Ref. \cite{12}) that
the reactor spectrum contains in addition to the Maxwell tail also the
spectrum of the moderated neutrons (Fermi spectrum). They averaged
$\ov{\hat\sig}_{\ga,Sm}(T_C,\De E_r)$  at  $T_C=600$K over this
spectrum and, comparing this curve with the experimental data of Fujii
et al. \cite{53}, found a shift $\De E_r=(-45^{+7}_{-15})$ meV. Thus
they found a shift of the cross section of $\Sm$. Let us consider the
\LT model in more detail in order to understand this result.
The age of the Oklo reactor is $T_0= 2\cdot10^9$ year.
The relative content of $\U$ is $\xi_5(0)=3.7\%$ (and not 3.1\%).
The ratio of $\text{H}$ to U is $f_H=N_H/N_U=3$.
The cross section of the burning up admixtures (e.g. lithium)
per atom of U is
$\be_U=\sum_i N_i\sig_a^i/N_U=2$ b.
The ratio of the  thermal neutron capture cross section
to the slowing down
cross section of epithermal neutrons is
$\De=\Sigma_a(V_T\bigl/(\Sigma_S/2A)=2$.
The temperature of the core is $T_C=600$K$=327^0$C.

The following
comments are appropriate concerning these parameters.
The age of the reactor is $1.8\cdot10^9$ year and not
$2\cdot10^9$ year.
The value of $\xi_5(0)=3.7\%$ yields cross sections
$\Sigma_{5,a}(0)$ and $\Sigma_{5,f}(0)$ which are too large by 1.2
times.
The ratio $f_H=N_H/N_U=3$ holds approximately for
$Y_{U3}(0)=59.6\%$ and $\om^0_{H_2O}=0.355$  (see Table \ref{tab3}).
For the condition $f_H=3$ to hold exactly one must reduce
$\om^0_{H_2O}=0.355$ to $\om^0_{H_2O}=0.323$.  The
concentration of uranium nuclei can be kept in the calculations at its
former value: $N_U=0.7205\cdot10^{-2}$ U/cm$\cdot$b.
The absorption cross section of $\mbox{}^{6}_{3}$Li nuclei per
uranium nucleus, $\beta=N_{Li}\sig_{a,Li}/N_U=2$ b at $T=300$K and
normal pressure results in the following concentration of lithium
nuclei: $N_{Li}=2.088\cdot10^{-4}$ Li/cm$\cdot$b.

The ratio of the capture cross section of thermal neutrons
to the scattering cross section of epithermal neutrons is too large in
the \LT paper. At $T=300$K, \LT use for the capture cross section the
value  $\sum_k \sig_{a,k}N_k/N_U=31.1$ b/U.
The scattering cross section of a free hydrogen nucleus is
20.5 b \cite{45} (for bound hydrogen it is greater). At $f_H=3$ the
value of $\De$ is $\De=2\cdot31.1/3\cdot20.5=1.01$ and not 2.  Even
though, in repeating the calculation of \LT we have used the value
$\De=2$. The curve $\hat\sig_{\ga,Sm}(T_C,\De E_r)$ is shown in Fig.
\ref{fig13} (curve {\em 1}).  The value of $\hat\sig_{\ga,Sm}^{\rm
Exp}(T)=(90.7\pm8.2)$ kb are taken from Table \ref{tab11}.
Recall that Table \ref{tab2} contains only four experimental
points instead of five. This results in a change of
$\hat\sig_{\ga,Sm}^{\rm Exp(T)}$  and its error as compared to \LT.
Curve
{\em 1} was obtained by interpolation of curves {\em 2} for $\De=1$
and for $\De=2$ to the value $\De=2\sqrt{300{\rm K}/600{\rm
K}}=\sqrt{2}$.  We have found a negative value for the energy
$\De E_{r2}=-24$ meV, closer to zero than the value
$\De E_{r2}=-38$ meV in \LT.  From
Fig.\ref{fig13} one can clearly see a specific feature of the result of
\LT.  It is enough to take $\De>1.41$ for the curve not to intersect the
error corridor, and for $\De<1.41$ the shift $\De E_{r2}$ is
strongly reduced.

\begin{figure}[h]
\epsfbox{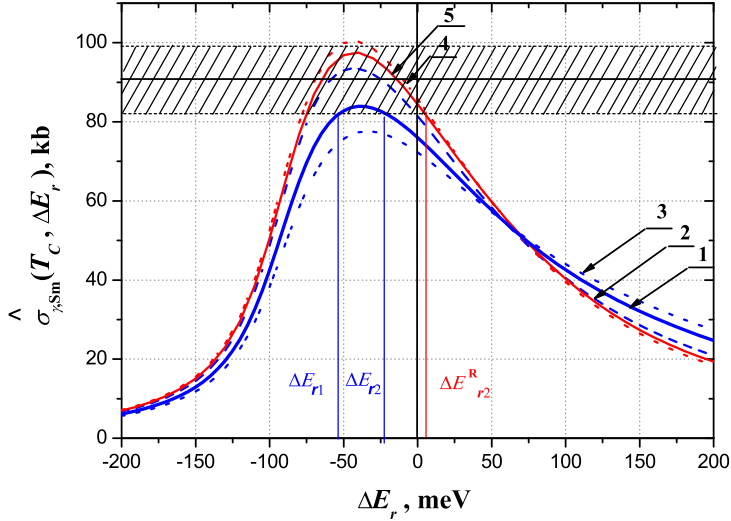}
\caption{\la{fig13}
Dependence $\hat\sig_{\ga Sm}(T,\De E_r)$ at $T_C=600$K for
the neutron flux of {\em Lamoreaux $\&$ Torgergerson} Ref. \cite{52},
calculated by us for $\De=\sqrt{2}$ ({\em curve 1}). For comparison we
show results for $\De=1$ ({\em curve  2}) and $\De=2$ ({\em curve 3}).
{\em Curves  4} and {\em 5} are our calculations for the core
composition of {\em LT}; {\em curve 4} is without the {\em power effect}
({\em PE}); {\em curve 5} with {\em PE} (computation with code
MCU-REA}
\end{figure}

At a stiffness parameter $\De\sim1$ the spectrum is distorted on
account of large absorption by the strong absorbers at small
energies.  Under these conditions the spectrum cannot be
considered to be Maxwellian. It can be found only by direct Monte Carlo
calculation.  We have carried out the calculation of the reactor
spectrum and of the distribution for a composition of the core as
described in items 3--5, using the code MCU REA. In the \LT paper no
absolute value of $N_U(0)$ was given, and we choose
$N_U=0.7205\cdot10^{-2}$ U/cm$\cdot$b.  The calculations were done both
without taking account of the {\em power effect} ({\em PE}) (curve {\em
4}) and with taking account of the {\em PE} (curve {\em 5}). Both
calculations cross the error corridor at $\De E_{r2}^R>0$ ($\De
E_{r2}^R=4$ meV). This means that in this case $\ad=0$ as well.

In the \LT model the neutron balance is
maintained on account of a compensation of the fuel burn-up by the
burn-up of strong absorbers.  This is far from reality.  Since strong
absorbers burn up faster than $\U$, such a balance exists only at the
beginning of the cycle. At the end of the cycle no strong absorber is
left. The fast burning up strong absorber $\mbox{}^{149}$Sm that is
still present to this day was formed only at the end of the cycle when
the \LT model does not work any more.

The results on a possible change of $\al$ based on the analysis of the
cross section of $\Sm$ in the Oklo reactor are summarized in Table
\ref{tab12}.  For comparison we have included the cosmological results
(Fig.\ref{fig14}) \cite{54a,54b} and the results of laboratory
measurements \cite{55}. A
review of a possible change of the fundamental constants (experiment
and theoretical interpretation) was recently published by Uzan
\cite{56a,56b}.  All results show that there are no grounds for an
assertion that the e.m.  constant has changed in the distant past.
However there is a possibility that this conclusion will be revised
when the fuel burn-up is taken into account.

\begin{figure}[h]
\epsfxsize=10cm
\epsfysize=10cm
\epsfbox{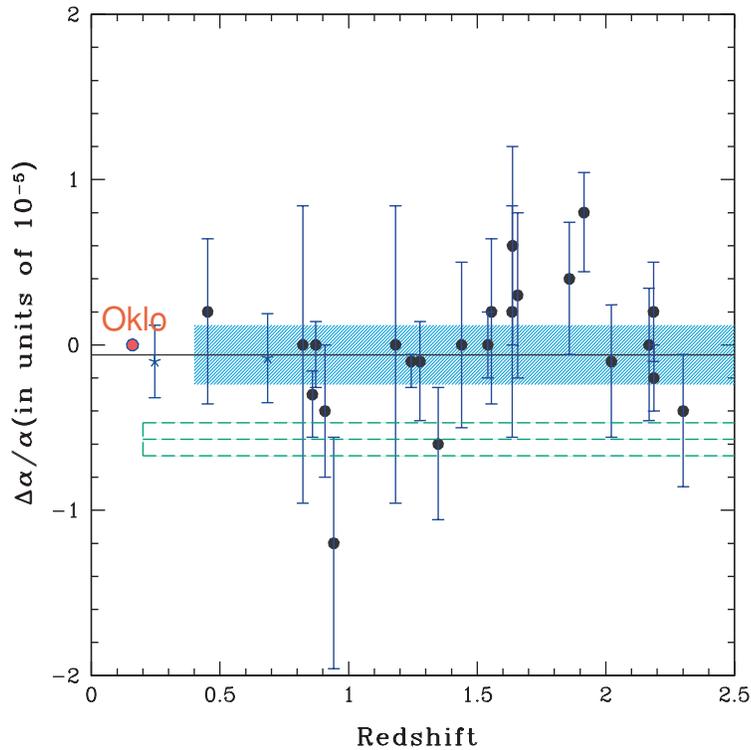}
{\caption{\la{fig14}
The data of {\em Chand, Srianand et al.} (filled circles) are plotted
against the redshift \cite{54b}. Each point is the best fitted value
obtained for individual systems using $\chi^2$ minimization. The open
cirle is measurement from the Oklo reactor.   The weighted average and
$1\sig$ range measured by Murphy et al. \cite{Murphy} are shown with
horizontal dashed lines. Most of Chand et al. measurements
are inconsistent with this range. The shaded region represents the
weighted average and its $3\sig$ error:
$<\de\al/\al>_W=(-60\pm60)\cdot10^{-8}$.  Within $3\sig$ there is no
variation of fundamental constants  within the limits:\\
$-25\cdot10^{-17}$ year$^{-1}\le(\De\al/\al\De t)\le 12\cdot10^{-17}$
year$^{-1}$.  }} \end{figure}

\begin{table} [h]
\caption{\la{tab12}
Limits on the rate of $\alpha$ variation,
based on the data of $\Sm$ contents in the Oklo
reactor\\
($z=1.8\cdot10^9$year$/13.7(2)\cdot10^9$year$=0.131(2)$);
 $t_0=13.7(2)$ \cite{PDG} and obtained by other methods.}

\begin{tabular}{|c|c|c|c|c|}
\hline
& Lab & Authors, year& $\De E_r$,
$\overline{\delta\alpha}/\al$, $\dot{\overline{\delta\alpha}}/\alpha$
& Comments \\
&&&& core, spectrum, $T_C$\\
\hline
1& LNPI, Gatchina & A.Shlyakhter,& $\De E_r\leq 20$ meV &
\RZ, Maxwell, 300 K\\
&Russia& 1976 \cite{6b} &
$\dot{\overline{\delta\alpha}}/\alpha\leq 0.5\cdot 10^{-17}$ year$\mbox{}^{-1}$ & \\
\hline
2& LNPI,& Yu. Petrov,& $\De E_r\leq 50$ meV & \RZ, Maxwell, 300K\\
& Gatchina,&1977  \cite{7} &
$\dot{\overline{\delta\alpha}}/\alpha\leq 2.5\cdot 10^{-17}$
year$\mbox{}^{-1}$ & \\
& Russia  & & &  \\
\hline
3&Princeton,& T.Damour and& $\De E_r\leq 90$ meV &
\RZ, Maxwell, (450-1000)K \\
& USA &F.Dyson, &
$\dot{\overline{\delta\alpha}}/\alpha\leq 5.0\cdot 10^{-17}$
year$\mbox{}^{-1}$ & \\
&   & 1996 \cite{8a,8b}& & \\
\hline
4& Univ.Tokio,& Ya.Fujii& $\De E_r\leq 20$ meV & {\em RZ10}, Maxwell,
(470-670)K\\
& Tokyo, & et al.,&
$\dot{\overline{\delta\alpha}}/\alpha\leq 1.0\cdot 10^{-17}$
year$\mbox{}^{-1}$ & \\
& Japan&  2000 \cite{9}& &\\
\hline
&&&&\\
5& LANL, & S.Lamoreaux and& $\De E_r\leq
-45^{\:\:+7}_{-15}$ meV & {\em RZ10}, Maxwell+Fermi,\\
& Los Alamos,& J.Torgerson,&
$\dot{\overline{\delta\alpha}}/\alpha\leq -3.8\cdot 10^{-17}$
year$\mbox{}^{-1}$ &
600 K
\\
& USA & 2004   \cite{54}& &\\
\cline{2-5}
&& This paper, & $\De E_r\leq 4$ meV  & {\em RZ10}, Reactor spectrum\\
&& Fig. \ref{fig13} &
$\dot{\overline{\delta\alpha}}/\alpha\leq 0.2\cdot 10^{-17}$
year$\mbox{}^{-1}$ &for \LT core,\\
&&&& 600 K\\
\hline
6& PNPI,& This paper & $\De E_r\leq 62$ meV & \RZ, Reactor spectrum\\
& Gatchina, &
&
$\dot{\overline{\delta\alpha}}/\alpha\leq 3.1\cdot 10^{-17}$
year$\mbox{}^{-1}$ &
for fresh core,\\
& Russia &&& $725\pm 55$ K\\
\hline
\multicolumn{5}{c}{}\\
\multicolumn{5}{c}{Cosmophysical and laboratory data}\\
\multicolumn{5}{c}{}\\
\hline
7 & IUCAA, Pune, & H.Chand, &
$\de\al/\al\leq(-60\pm60)\cdot10^{-8}$ &
Cosmophysical\\
&India
& R.Srianand et al., & $\dot{\overline{\delta\alpha}}/\alpha\leq 12
\cdot 10^{-17}$ year$\mbox{}^{-1}$ &
multidoublet method\\
&& 2004 \cite{54a,54b} &&\\
\hline
8& Observ.&S.Bize et al.& $\dot{\ov{\de\al}}/\al\leq(-5\pm53)\cdot10^{-17}$
year$^{-1}$& Method of \\
& de Paris, & 2004 \cite{55} & &   Atomic
Fountains \\
& France &&& \\
\hline
\end{tabular}
\end{table}

\noindent

\section{Conclusions}

We have built a complete computer model of the Oklo reactor core
\RZ. With the aid of present-day computational codes we have
calculated in all detail the core parameters. The simulations
were done for three fresh cores of different contents of
uranium and water. We have also calculated the neutron flux and its
spatial and energy distributions. For the three cores we have
estimated the {\em temperature} and {\em void effects} in the reactor.
As expected, the neutron reactor spectrum is significantly
different from the ideal Maxwell distribution that had been used by
other authors to determine the cross section of $\Sm$.
The reactor cross section and the curves of its dependence on
the shift of the resonance position $\De E_r$  (as a result of a
possible change of fundamental constants) differ appreciably from
earlier results. {\em The effect of an influence of the reactor
spectrum} on the cross section of $\Sm$ can be considered to be firmly
established. We have studied the limits of the variation of this effect
depending on the initial composition and the size of the core.
The fresh bare core \RZ  is critical for
$T_C=(725\pm55)$K. At these temperatures the curves of
$\hat\sig_{\ga,Sm}(T_C,\De E_r)$ lie
appreciably lower than for a Maxwell distribution.  Possible values of
$\De E_r$ lie in the range of $-73$ meV $\leq\De E_r\leq62$ meV. These
limits are 1.5 times more accurate than those of Dyson and Damour. For
the rate of change of the e.m.  constant we find $-3.7\cdot10^{-17}$
year$^{-1} \leq\ad\leq +3.1\cdot10^{-17}$ year$^{-1}$. Within
these limits we have $\ad=0$.
The
analysis of all previous studies shows that none of them has reliably
shown up a deviation from zero of the rate of change of the e.m.
constant $\al$.
Because of difficulties with the
detailed calculation of the burning up in large reactors, which require
accumulation of huge statistics, we have not determined the effect of
the burn-up on the neutron spectrum and on the $\Sm$ cross section.
Calculations of the influence of burn-up on the temperature of the
active core and on the neutron spectrum are in progress.

\begin{acknowledgments}
The authors express their thanks to V.A. Varshalovich and
B.L. Ioffe for discussions, also their appreciation to
N.N. Ponomarev-Stepnoi, E.A. Gomin, M.I. Gurevich,
A.S. Kalugin and M.S. Yudkevich for making available
codes MCU-REA and BURNUP.
The authors consider it their pleasant duty to
thank V.V. Kuzminov for making available the
nuclear constants for code MCNP4C, J. Vallenius for consultations and
W.B. von Schlippe for the translation.
\samepage
This work was done with the partial financial support of grant
RFFI 02-02-16546-a.
\end{acknowledgments}



\end{document}